\documentclass[espf]{aa}
\usepackage{graphicx}
\usepackage{amsmath}
\usepackage[psamsfonts]{euscript}
\usepackage[psamsfonts]{amssymb}
\usepackage{amsmath}
\usepackage{graphicx}
\usepackage{natbib}
\usepackage{epsfig}

\titlerunning{}

\begin{document}
\title{High resolution spectroscopy for Cepheids distance
determination}


\subtitle{I. Line asymmetry}
\titlerunning{High resolution spectroscopy for Cepheids distance
determination}
\authorrunning{N. Nardetto et al.}

\author{ N. Nardetto \inst{1}, D. Mourard \inst{1}, P. Kervella \inst{2}, Ph. Mathias \inst{1}, A. M\'erand \inst{2}, D. Bersier
\inst{3,4}}

\institute{Observatoire de la C\^ote d'Azur, Dpt. Gemini, UMR 6203,
F-06130 Grasse, France %
\and Observatoire de Paris-Meudon, LESIA, UMR
8109, 5 Place Jules Janssen, F-92195 Meudon Cedex, France %
\and Space Telescope Science Institute, 3700 San Martin Drive,
Baltimore, MD
21218, USA %
\and Astrophysics Research Institute, Liverpool John Moores
University, Twelve Quays House, Egerton Wharf, Birkenhead,
CH41 1LD, UK %
}

\date{Received ... ; accepted ...}

\abstract {The ratio of pulsation to radial velocity (the projection
factor) is currently limiting the accuracy of the Baade-Wesselink
method, and in particular of its interferometric version recently
applied to several nearby Cepheids.} {This work aims at establishing
a link between the line asymmetry evolution over the Cepheids'
pulsation cycles and their projection factor, with the final
objective to improve the accuracy of the Baade-Wesselink method for
distance determinations.} {We present HARPS\thanks{High Accuracy
Radial velocity Planetary Search project developed by the European
Southern Observatory} high spectral resolution observations
($R=120000$) of nine galactic Cepheids : \object{R~Tra},
\object{S~Cru}, \object{Y~Sgr}, \object{$\beta$~Dor},
\object{$\zeta$~Gem}, \object{Y~Oph}, \object{RZ~Vel},
\object{$\ell$~Car} and \object{RS~Pup}, having a good period
sampling ($P=3.39$d to $P=41.52$d). We fit spectral line profiles by
an asymmetric bi-gaussian to derive radial velocity, Full-Width at
Half-Maximum in the line (FWHM) and line asymmetry for all stars. We
then extract correlations curves between radial velocity and
asymmetry. A geometric model providing synthetic spectral lines,
including limb-darkening, a constant FWHM (hereafter $
\sigma_{\mathrm C}$) and the rotation velocity is used to interpret
these correlations curves.} {For all stars, comparison between
observations and modelling is satisfactory, and we were able to
determine the projected rotation velocities and $ \sigma_{\mathrm
C}$ for all stars. We also find a correlation between the rotation
velocity ($V_{\mathrm rot} \sin i$) and the period of the star:
$V_{\mathrm rot} \sin i= (-11.5 \pm 0.9) \log(P) + (19.8 \pm 1.0)
\mbox{[km.s$^{-1}$]}$. Moreover, we observe a systematic shift in
observational asymmetry curves (noted $\gamma_{\mathrm O}$), related
to the period of the star, which is not explained by our static
model : $\gamma_{O}=(-10.7 \pm 0.1) \log(P) + (9.7 \pm 0.2) \mbox{
[in \%] }$. For long-period Cepheids, in which velocity gradients,
compression or shock waves seem to be large compared to short- or
medium-period Cepheids we observe indeed a greater systematic shift
in asymmetry curves.} {This new way of studying line asymmetry seems
to be very promising for a better understanding of Cepheids
atmosphere and to determine, for each star, a dynamic projection
factor.}


\maketitle

\section{Introduction}\label{s_Introduction}

\begin{table*}
\begin{center}
\caption[]{Observed sample of Cepheids sorted by increasing period.
\label{Tab_Cepheids_Parameters}}
\begin{tabular}{lcccccccccc}
\hline \hline \noalign{\smallskip}

Name     &  HD     &  $P$~{\tiny (a)}          & $T_0$ {\tiny (a)}                     & Nb.               & Nb. of            & ${m_\mathrm{V}}$ {\tiny (b)} \\
         &         &  [days]         & [days]                                &  of spectra       & cycles                                             \\
\hline
R TrA       & 135592  &   $3.38925$     &  $2\,451\,649.96$         &       14               &               15 &      6.66\\
S Cru       & 112044  &   $4.68976$     &  $2\,451\,645.64$         &       12               &               3  &  6.60   \\
Y Sgr       & 168608  &   $5.77338$     &  $2\,451\,650.92$         &       17               &               10 & 5.74  \\
$\beta$ Dor & 37350   &   $9.84262$     &  $2\,451\,643.54$         &       49               &            3     & 3.75   \\
$\zeta$ Gem & 52973   &   $10.14960$    &  $2\,451\,641.78$         & 50                     & 3                & 3.90   \\
Y Oph       & 162714  &   $17.12520$    &  $2\,451\,653.32$         & 7                      & 4                & 6.17  \\
RZ Vel      & 73502   &   $20.40020$    &  $2\,451\,633.58$         & 10                     &  3               & 7.08  \\
$\ell$~Car  & 84810  &   $35.551341$    &  $2\,452\,290.4158$          & 118                    & 2                & 3.74  \\
RS Pup      & 68860   &   $41.51500$    &  $2\,451\,644.22$         & 15                     & 3                & 7.03 \\

\hline \noalign{\smallskip}
\end{tabular}
\end{center}
\begin{list}{}{}

\item[$^{\mathrm{a}}$] For $\ell$~Car, the reference Julian date ($T_0$) and the pulsation
period ($P$) used to compute the phase are from Szabados
\cite{Szabados89}. For others stars we used ephemeris from Berdnikov
et al. \cite{Berdnikov01}.

\item[$^{\mathrm{b}}$] The visible magnitude (${m_\mathrm{V}}$) is
from Berdnikov et al. \cite{Berdnikov00}.

\end{list}

\end{table*}

Long-baseline interferometers currently provide a new
quasi-geometric way to calibrate the Cepheid Period-Luminosity
relation. Indeed, it is now possible to determine the distance of
galactic Cepheids up to $1$kpc with the Interferometric
Baade-Wesselink method, hereafter IBW method (see for e.g. Sasselov
\& Karovska \cite{Sasselov94} and Kervella et al. \cite{Kervella04},
hereafter Paper I). Interferometric measurements lead to angular
diameter estimations over the whole pulsation period, while the
stellar radius variations can be deduced from the integration of the
pulsation velocity. The latter is linked to the observational
velocity deduced from line profiles by the projection factor $p$. In
this method, angular and linear diameters have to correspond to the
same layer in the star to provide a correct estimate of the
distance.

The spectral line profile, in particular its asymmetry, is
critically affected by the dynamical structure of Cepheids'
atmosphere : photospheric pulsation velocity (hereafter $V_{puls}$),
velocity gradients, limb-darkening, turbulence and rotation. Thus,
radial velocities measured from line profiles, hereafter $V_{rad}$,
include the integration in two directions : over the surface,
through limb-darkening, and over the radius, through velocity
gradients. All these phenomena, except the rotation, are supposed to
vary with the pulsation phase. However, they are currently merged in
one specific quantity, generally considered as constant with time:
the projection factor $p$, defined as $V_{puls}=pV_{rad}$.

The interferometric definition of the projection factor is of
crucial importance in the IBW method, as it can induce a bias of up
to 6\% on the derived distance (Nardetto at al. \cite{Nardetto04},
M\'erand et al. \cite{Merand05}). Otherwise, the limb-darkening is
also required to derive a correct estimation of the angular diameter
of the star. With the latest generation of long-baseline
interferometers, studying its phase-dependence is of crucial
importance (Marengo et al. \cite{Marengo02}, \cite{Marengo03},
Nardetto et al. \cite{Nardetto06}).

Line asymmetry was first observed for short-period cepheids by
Sasselov et al. \cite{Sasselov89}. Then, Sasselov et al.
\cite{Sasselov90} studied the impact of the asymmetry on radius and
distances determinations. The link between line profiles asymmetry
and the projection factor has been studied by Albrow et al.
\cite{Albrow94}. Finally, an error analysis of the IBW method is
given in Marengo et al. \cite{Marengo04}.

We present here a new original study of the line asymmetry using the
very high spectral resolution of HARPS ($R=120000$). We have
observed 9 galactic Cepheids with periods ranging from P=3.39\,$d$
to P=41.52\,$d$. Radial velocity, full-width at half-maximum
(hereafter FWHM) and line asymmetry are presented for all stars in
Sect. \ref{s_Observations}.

Section \ref{s_modelling} deals with modelling and Sect.
\ref{s_interpretation} with observations interpretation. Through a
geometric model different definitions of the projection factor are
proposed and compared in order to find the best procedure. Then the
model is used to interpret observational radial velocity and
asymmetry correlation curves. A set of parameters is thus derived
for all stars. Taking into account the whole sample of stars we
discuss general properties and in particular the
period-dependencies.

\section{HARPS observations} \label{s_Observations}

\subsection{Journal of observations}\label{ss_Data_reduction}

HARPS is a spectrometer dedicated to the search for extrasolar
planets by means of radial velocity measurements. It is installed at
the Coud\'e room of the $3.6$ meter telescope at La Silla. The
resolution is $R=120000$ and the average Signal to Noise Ratio we
obtain over all observations in the continuum (292 spectra) is $300$
per pixel. The observed sample of Cepheids is presented in
Table~\ref{Tab_Cepheids_Parameters}.

We have used the standard ESO/HARPS pipe-line reduction package with
a special attention for the normalization process. We have noted on
metallic line profiles of all stars a good reproduction from
cycle-to-cycle. Therefore, spectra for a given star have been
recomposed into an unique cycle.

Using Kurucz models \cite{kurucz92} we have identified about 150
unblended spectral lines. This first study considers only the
unblended metallic line \ion{Fe}{I} 6056.005 \AA.

\subsection{A new estimator of the radial velocity, FWHM and asymmetry: the
bi-gaussian}\label{ss_BIestimator}
 Several methods have been used to measure radial velocities of
Cepheids, each having advantages and drawbacks.  Among these methods
there is the line minimum (usually determined via a parabolic fit to
a few pixels near the bottom of the line) a gaussian fit (obviously
not adequate for asymmetric lines), the line centroid, determined
from the integration of the line profile (requires high
$Signal/Noise$ ratio), and the line bisector where one measures the
width of the line at one or several depths. Our bi-gaussian approach
combines advantages of methods useful for low $S/N$ data while
providing information usually associated with high resolution and
high $S/N$ data (asymmetry).

Radial velocity, full width at half-maximum (FWHM) and asymmetry
have been derived simultaneously applying a classical $\chi^{2}$
minimization algorithm between the observed line profile
($S(\lambda)$) and a modelled spectral line profile ($f(\lambda)$).
The corresponding reduced $\chi^2$ is :

\begin{equation} \label{Eq_X2_BiGaussian}
{\chi_{\mathrm red}}^{2} = \frac{1}{N-\nu}
\sum_{i=0}^{N}{\frac{(S(\lambda_i)-f(\lambda_i))^{2}}{\sigma(\lambda_i)^2}}
\end{equation}

with $N$ the number of pixel in the spectral line, $\nu$ the number
of degrees of freedom and $\sigma(\lambda_i)=$ {\it SNR}
$*f(\lambda_i)$ is the statistical uncertainty associated to each
pixel. {\it SNR} is the estimate of the Signal to Noise Ratio in the
continuum.

The analytic line profile is defined by :

\begin{equation} \label{Eq_BiGaussian1}
f(\lambda)= 1 - D \exp \left( \frac{4 \ln 2 (\lambda -
\lambda_{\mathrm m}) ^2}{(FWHM(1+A))^2} \right) \mbox{ if } \lambda
> \lambda_{\mathrm{m}}
\end{equation}

and

\begin{equation} \label{Eq_BiGaussian2}
 f(\lambda)= 1  - D \exp \left( \frac{4 \ln 2 (\lambda - \lambda_{\mathrm m})
^2}{(FWHM(1-A))^2} \right) \mbox{ if } \lambda <
\lambda_{\mathrm{m}}
\end{equation}

with four free parameters :

\begin{itemize}

\item $D$, the depth of the line. This quantity has no dimension.

\item $\lambda_{\mathrm m}$, the wavelength associated to the
minimum of the line (in \AA) . The corresponding radial velocity is
noted $RV_{\mathrm{m}}$.

\item {\it FWHM} is the Full-Width at Half-Maximum in the line,
also in \AA.

\item $A$ is the asymmetry as a percentage of the $FWHM$.

\end{itemize}

The $4 \ln 2$ factor is to obtain a correct definition of the
$FWHM$. Forcing asymmetry to zero in this minimization process is
equivalent to fitting a gaussian to the line profile. In this case
we can derive another type of radial velocity noted $RV_{\mathrm
g}$.

There are different ways to define the line asymmetry (see e.g.
Sasselov et al. 1990; Sabbey et al. 1995). The advantage of the
bi-gaussian method is that it offers the possibility to derive
statistical uncertainties directly from the minimization process.
Moreover, all parameters ($RV_{\mathrm m}$, {\it FWHM}, {\it D} and
{\it A}) are fitted simultaneously leading to a very consistent set
of information. The largest reduced $\chi^2$ we obtain with this
method is of about $10$ corresponding to a $SNR$ of $438$, but in
most cases we have a reduced $\chi^2 \simeq 1$ or $2$ corresponding
to a $SNR$ ranging from $75$ to $350$. That means that our analytic
model is well suited to the data quality. We note also that the
reduced $\chi^2$ is not sensitive to the spectral line resolution.

\begin{figure}[]
\resizebox{\hsize}{!}{\includegraphics[clip=true]{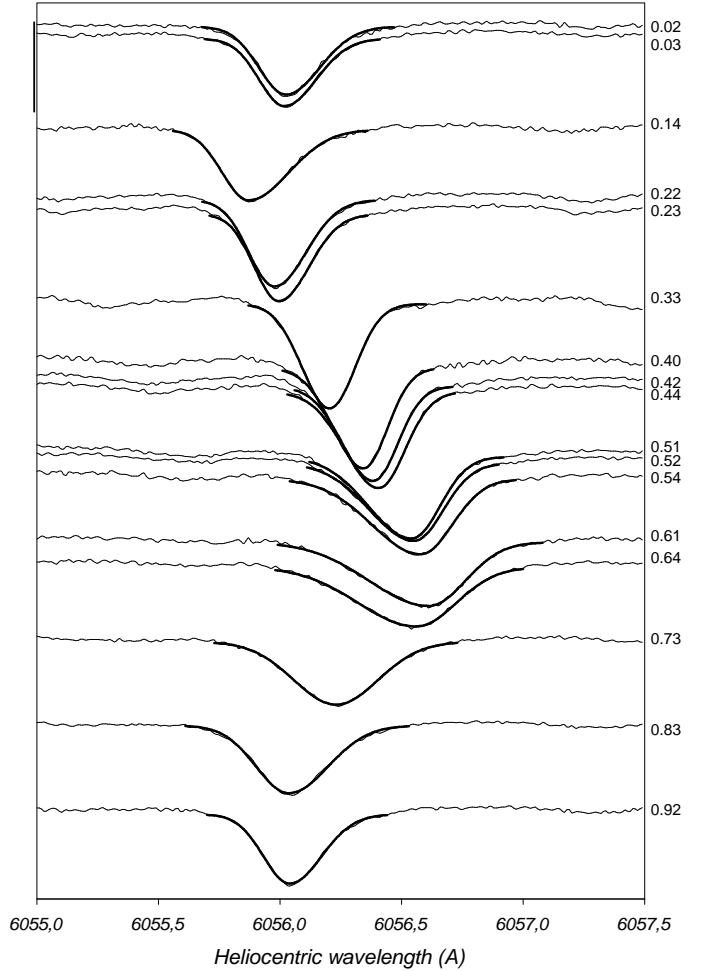}}
\caption{Spectral line evolution of $\beta$ Dor together with the
modelled bi-gaussian (bold). Line asymmetry is clear. The vertical
line at the top corresponds to a differential flux of $0.3$.
Pulsation phases are given on the right of each profile.}
\label{Fig_BDFilmp}
\end{figure}

As an example, Fig. \ref{Fig_BDFilmp} presents line profile
variation for $\beta$ Dor together with the analytic spectral line
profile. We find that the asymmetry is insensitive to the choice of
the continuum. However, this one has to be correctly defined to
derive correct values of the $FWHM$ and line depth $D$.

Another radial velocity definition, the centroid velocity
($RV_{\mathrm c}$) or, the first moment of the spectral line
profile, has been estimated as:

\begin{equation} \label{Eq_CDG}
 RV_{\mathrm c} = \frac{\int_{\rm line} \lambda S(\lambda) d\lambda}{\int_{\rm line} S(\lambda) d\lambda}
\end{equation}


Tables \ref{Tab_cepheids_results1}, \ref{Tab_cepheids_results2} and
\ref{Tab_cepheids_results3} present the resulting values of
$RV_{\mathrm{g}}$, $RV_{\mathrm{m}}$, $RV_{\mathrm{c}}$, $FWHM$,
$D$, $A$, $SNR$ and $\chi_{red}^{2}$ together with the corresponding
uncertainties computed from the fitting method.

\subsection{Radial velocity}\label{ss_RV}

\begin{figure}[]
\resizebox{\hsize}{!}{\includegraphics[clip=true]{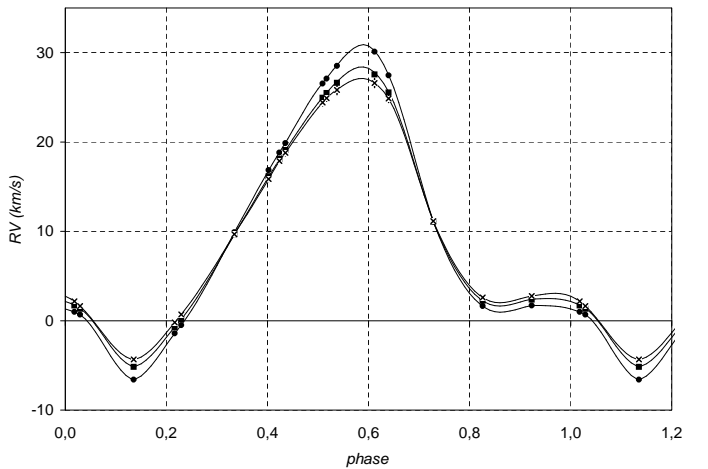}}
\caption{$\beta$ Dor radial velocities obtained with different
 method: $RV_{\mathrm m}$ (points), $RV_{\mathrm
g}$ (squares), and $RV_{\mathrm c}$ (crosses). Statistical
uncertainties at $\pm 1 \sigma$ are indicated but too small to be
visualized. We can therefore see the impact of the choice of the
method in the case of a very asymmetric line (Fig.
\ref{Fig_BDFilmp}).
 \label{Fig_BD_RV}}
\end{figure}

\begin{figure}[]
\resizebox{\hsize}{!}{\includegraphics[clip=true]{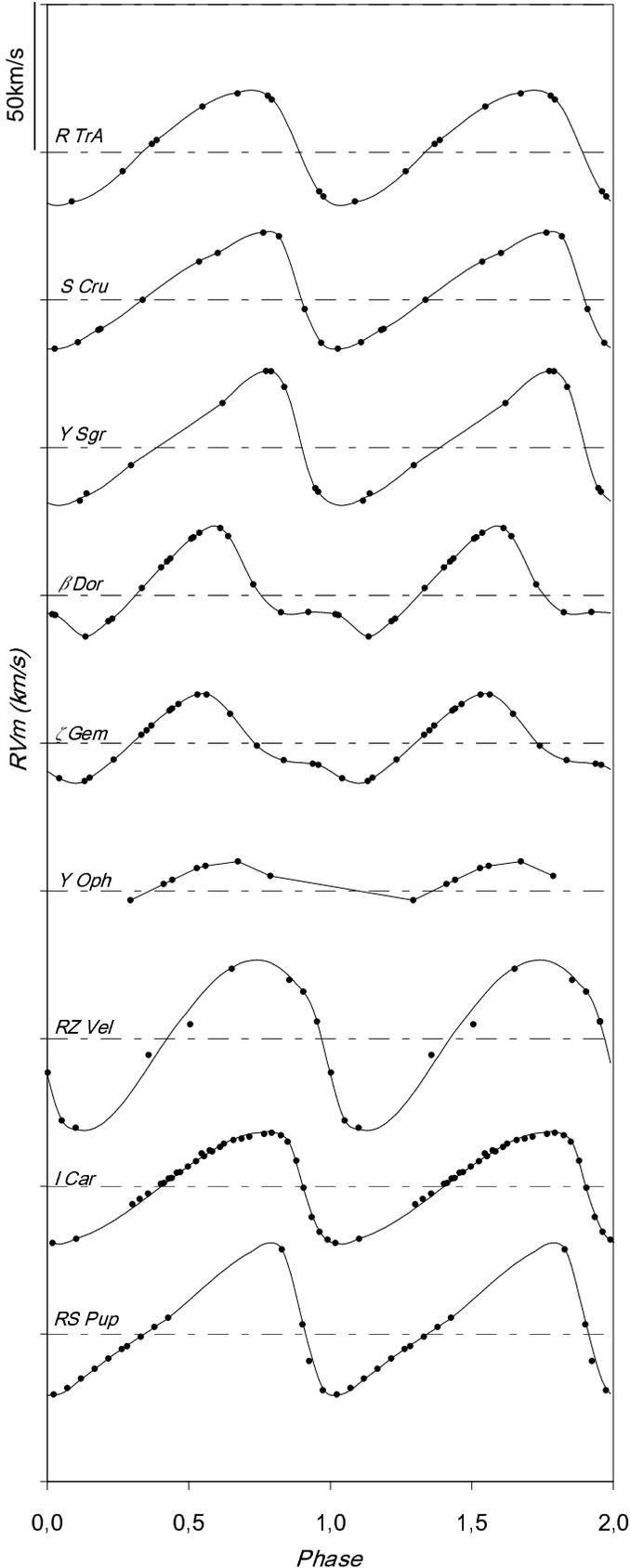}}
\caption{Radial velocity curves ($RV_{\mathrm m}$). Curves have been
arbitrarily shifted vertically. The horizontal lines are the zero
velocity in the stellar rest frame. Largest velocities are for
receding motion. }


\label{Fig_RVmShifted}

\end{figure}

As indicated in the previous section, we can derive three types of
radial velocity : the velocity associated to the gaussian fit
($RV_{\mathrm g}$), the line minimum  ($RV_{\mathrm{m}}$) and the
barycenter of the spectral line ($RV_{\mathrm c}$). Figure
\ref{Fig_BD_RV} shows these radial velocity curves obtained in the
case of $\beta$ Dor. Figure \ref{Fig_RVmShifted} represents for each
star of our sample, the $RV_{\mathrm m}$ variation (arbitrary
shifted). The solid lines are the interpolated curves using a
periodic cubic spline function. This function is calculated either
directly on the observational points (e.g. $\beta$~Dor) or using
arbitrary pivot points (e.g. RZ~Vel). In the latter case, a
classical minimization process between observations and the
interpolated curve is used to optimize the position of the pivot
points. All the interpolated curves presented in this study are
derived using one of these two methods. The only exception is Y~Oph
(too few points)  for which we performed a linear interpolation.

\subsection{The Full-Width at Half-Maximum in the line}\label{ss_FWHM}

\begin{figure}[]
\resizebox{\hsize}{!}{\includegraphics[clip=true]{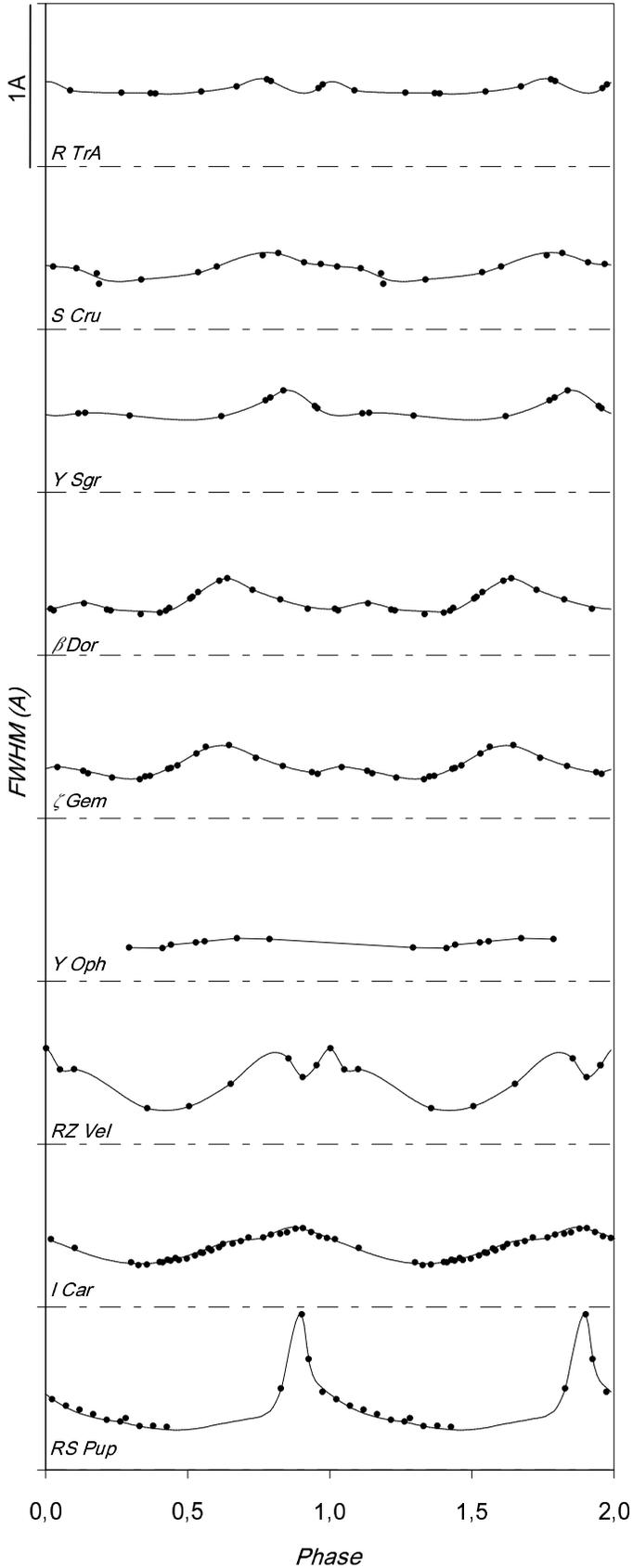}}
\caption{FWHM versus phase for all stars. Curves have been
arbitrarily shifted vertically. The horizontal lines correspond to a
zero FWHM. Note the particular case of RS Pup, which may present the
signature of an important compression or shock wave. RS Pup has the
longest period of our sample.} \label{Fig_FWHM}
\end{figure}

\begin{figure}[]
\resizebox{\hsize}{!}{\includegraphics[clip=true]{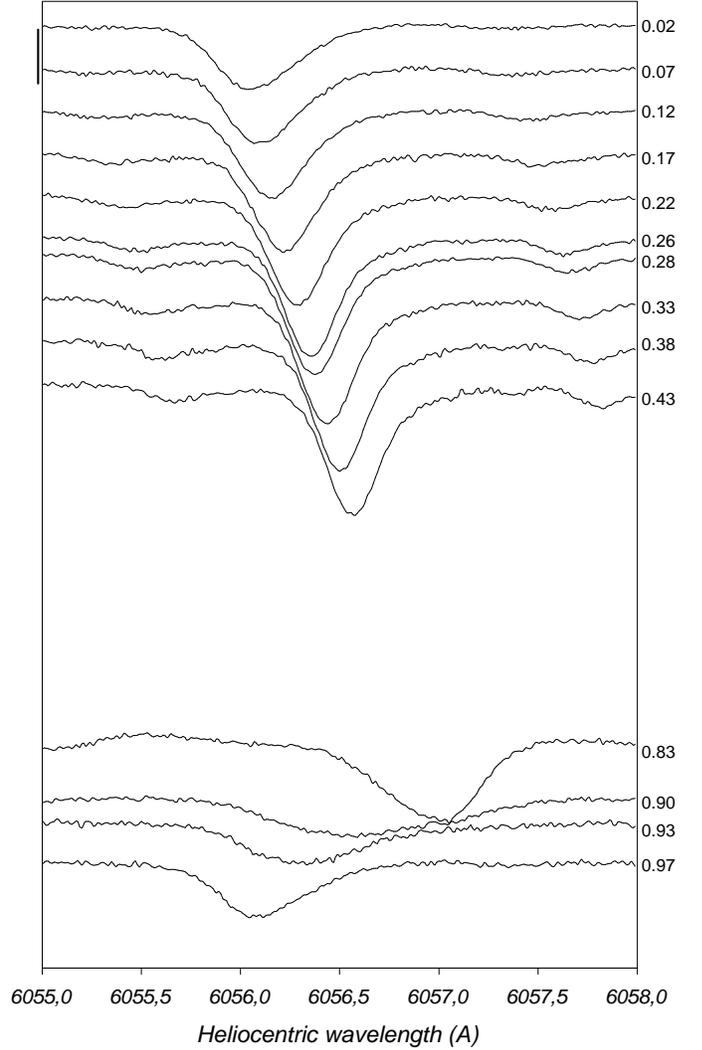}}
\caption{FeI 6056.005 \AA\ spectral line evolution of RS~Pup. The
vertical line at the top corresponds to a differential flux of
$0.2$. We note the broadening of the line at $\phi=0.83$ which could
be the signature of a strong velocity gradient (compression or shock
wave).} \label{Fig_RSPFilmp}
\end{figure}

Figure \ref{Fig_FWHM}  presents the FWHM curve as a function of
phase for all stars. We note that the largest FWHM values are
obtained for the maximum contraction velocities. RS~Pup, the longest
period Cepheid of our sample, seems to present an important
compression or shock wave signature. Figure~\ref{Fig_RSPFilmp}
presents line profile variation for this star. Unfortunately the
phase coverage is not very good, but we can clearly see a strong
increase of the {\it FWHM} at $\phi=0.83$. Such phenomenon has been
already detected in $\beta$ Cepheids (Fokin et al. \cite{fokin04}).

\subsection{Asymmetry}\label{ss_Asymmetry}

\begin{figure}[]
\resizebox{\hsize}{!}{\includegraphics[clip=true]{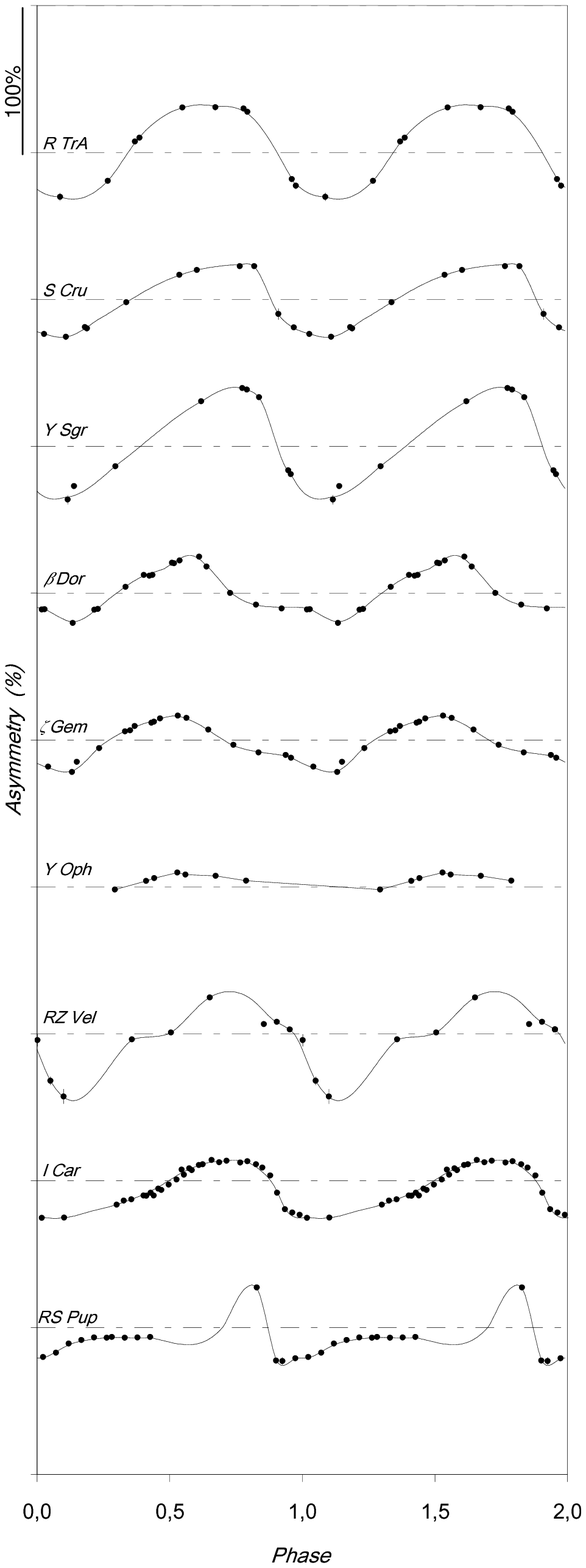}}
\caption{Asymmetry against phase for all stars. Curves have been
arbitrarily shifted vertically. The horizontal lines correspond to
an asymmetry of zero. } \label{Fig_ASY}
\end{figure}

Figure \ref{Fig_ASY} shows the asymmetry variation for all stars.
Generally speaking, the shape of the aymmetry curve is similar to
the shape of the velocity curve $RV_m$.

\begin{figure}[]
\resizebox{\hsize}{!}{\includegraphics[clip=true]{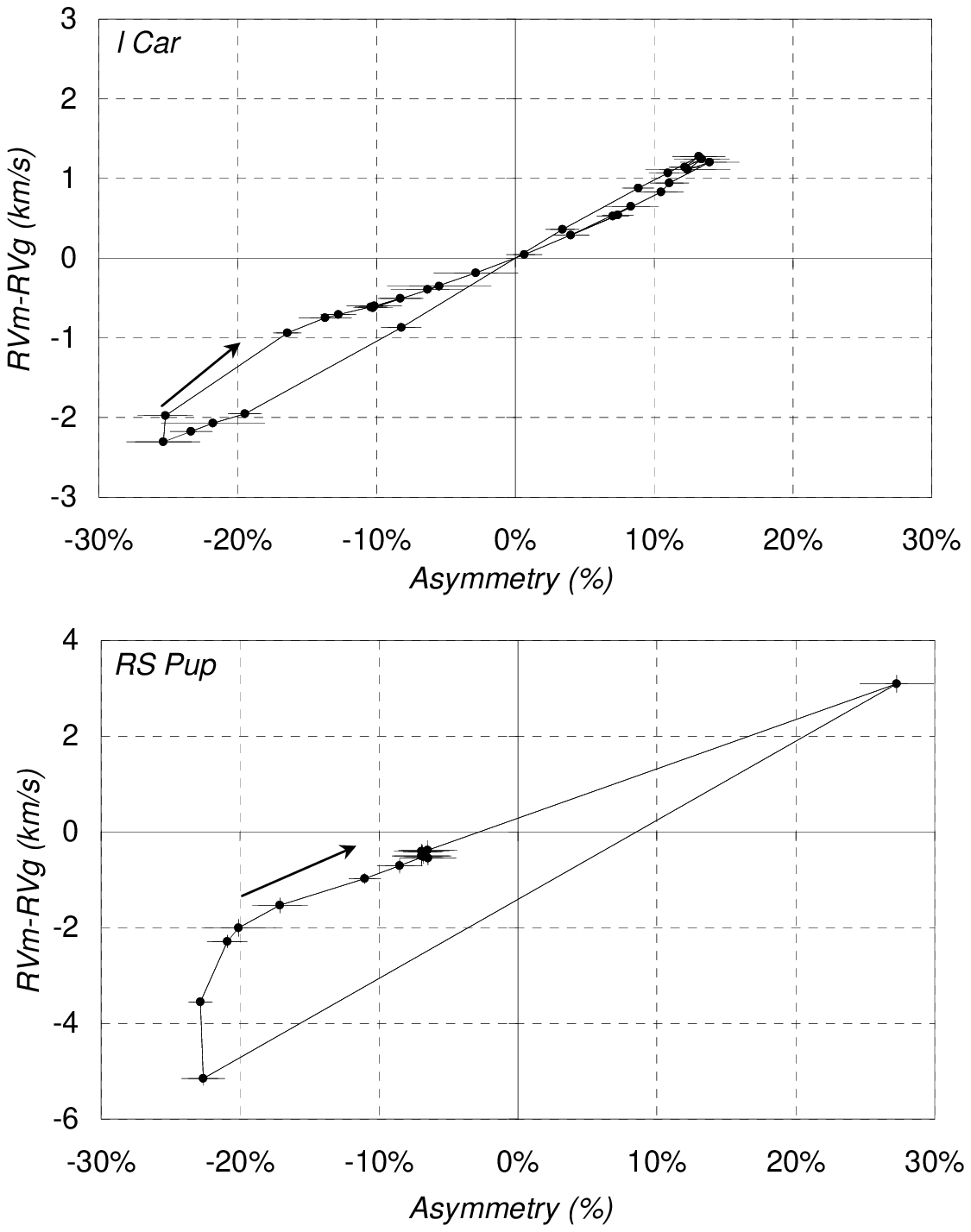}}
\caption{ Difference between the radial velocity obtained with the
line minimum and the gaussian fit methods as a function of the
asymmetry in the case of $\ell$~Car and RS~Pup. Statistical
uncertainties are provided for each point. Arrows indicate the
direction and the origin $\phi=0$ of the curves. These relations are
not linear and certainly affected by star characteristics (rotation,
FWHM, velocity gradients...). } \label{Fig_Correlation_Erreurs_Asy}
\end{figure}

As already mentioned in Sect. \ref{ss_RV}, the radial velocity
according to the choice of the method considered is sensitive to the
line asymmetry. Figure \ref{Fig_Correlation_Erreurs_Asy} shows the
correlation between the differences of radial velocity ($\Delta V =
RV_{\mathrm m}-RV_{\mathrm g}$) and the asymmetry of the line. We
have only presented here the case of $\ell$~Car and RS~Pup. Each
star presents a similar behavior. A typical difference in velocity
of about $4$km.s$^{-1}$ can be obtained for an asymmetry of $40\%$
in extreme cases (Y~Sgr and R~TrA). The relation between the radial
velocity difference and the asymmetry is certainly affected by star
characteristics (rotation, FWHM, velocity gradients) present in the
line asymmetry. In particular RS~Pup signature is certainly affected
by strong velocity gradient effects. The fact that the $RV_{\mathrm
m}$ and $RV_{\mathrm g}$ radial velocities present such differences
as a function of the pulsation phase is an additional difficulty
concerning an average projection factor and its time-dependence
determination. With the centroid estimator of the radial velocity
($RV_{\mathrm g} - RV_{\mathrm c}$ or $RV_{\mathrm m} - RV_{\mathrm
c}$) results are quite similar.

In next sections, we summarize all observational results in
correlation diagrams between radial velocity and asymmetry. These
correlations are interpreted using the geometric model in order to
determine some physical parameters of our stars and to obtain
information about dynamical effects in Cepheids atmosphere.

\section{A toy model} \label{s_modelling}

We consider a limb-darkened pulsating star in rotation with an
one-layer atmosphere. Our model has four parameters :

\begin{itemize}
\item the limb-darkening of the star : we consider a
linear law for the continuum-intensity profile of the star defined
by $I(\cos(\theta))=1-u_{\mathrm V}+u_{\mathrm V}\cos(\theta)$,
where $u_{\mathrm V}$ is the limb-darkening of the star in V band
(Claret et al. \cite{claret00}). Its value is about $0.7$ for
Cepheids. $\theta$ is the angle between the normal of the star and
the line-of-sight.

\item the projected rotation velocity $V_{\mathrm{rot}} \sin
i$, where $i$ is the angle between the line-of-sight and the
rotation axis (in km.s$^{-1}$).

\item the pulsation velocity (in km.s$^{-1}$).

\item the width of the spectral line (in \AA), hereafter named $ \sigma_{\mathrm C}$. It is
the FWHM of the line with no pulsation nor rotation velocities. It
is supposed to be constant with the pulsation phase.

\end{itemize}

\begin{figure}[]
\resizebox{\hsize}{!}{\includegraphics[clip=true]{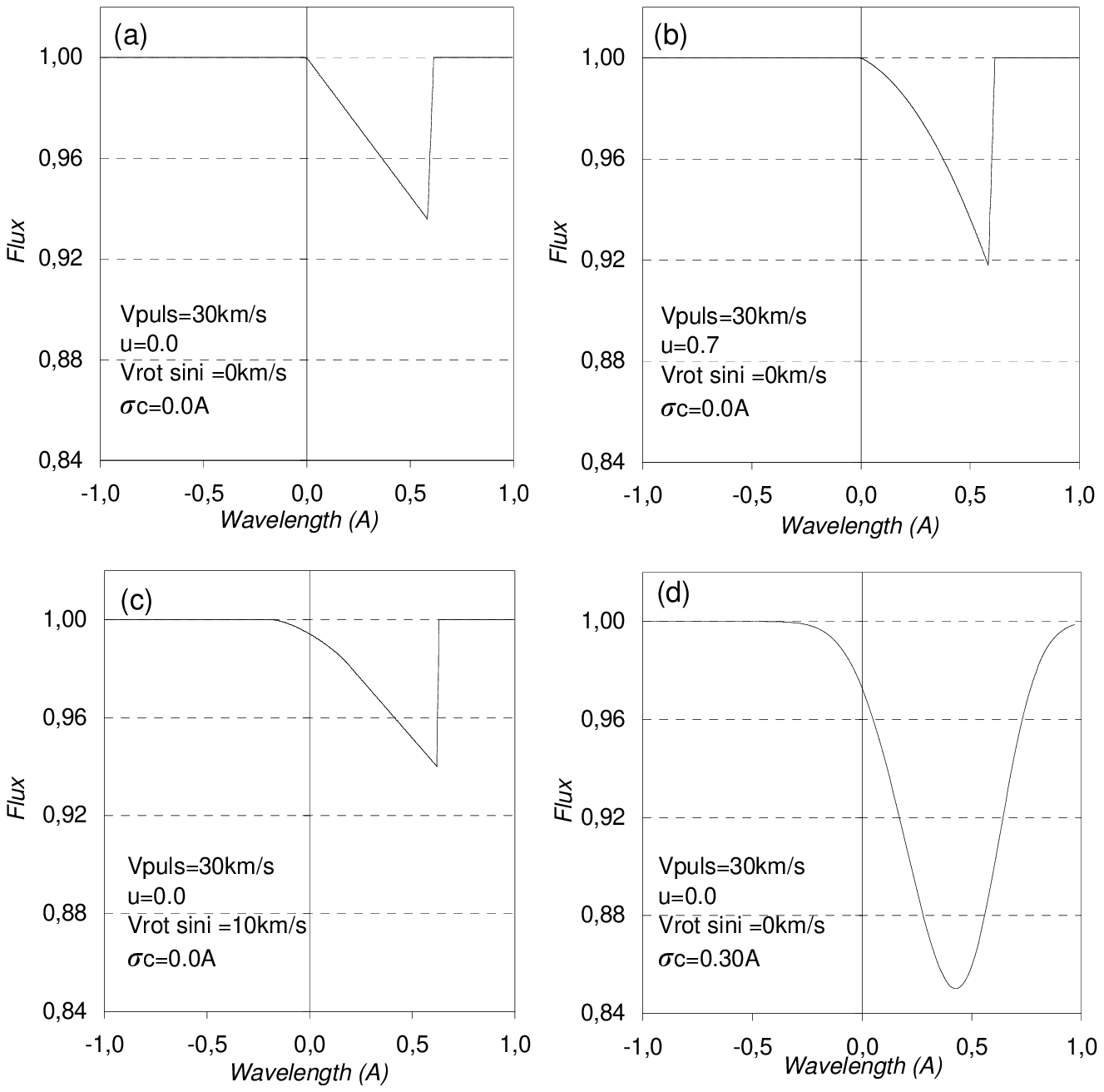}}
\caption{The weighting or the synthetic spectral line profile in
different cases, considering (a) the pulsation velocity, (b) the
limb-darkening, (c) the rotation and, (d) an intrinsic width for the
line ($\sigma_{\mathrm C}$). } \label{Fig_model}
\end{figure}

The velocity field is a combination of pulsation and rotation
velocities. Through the Doppler effect, this field can be transposed
into wavelengths, and weighted by the surface brightness
(limb-darkening) to obtain the weighting of the spectral line. We
have then to convolve it with the intrinsic profile to obtain the
synthetic spectral line profile. The weighting or the synthetic
spectral line profile are presented in different cases in Fig.~
\ref{Fig_model}.


\begin{figure*}[]
\resizebox{\hsize}{!}{\includegraphics[clip=true]{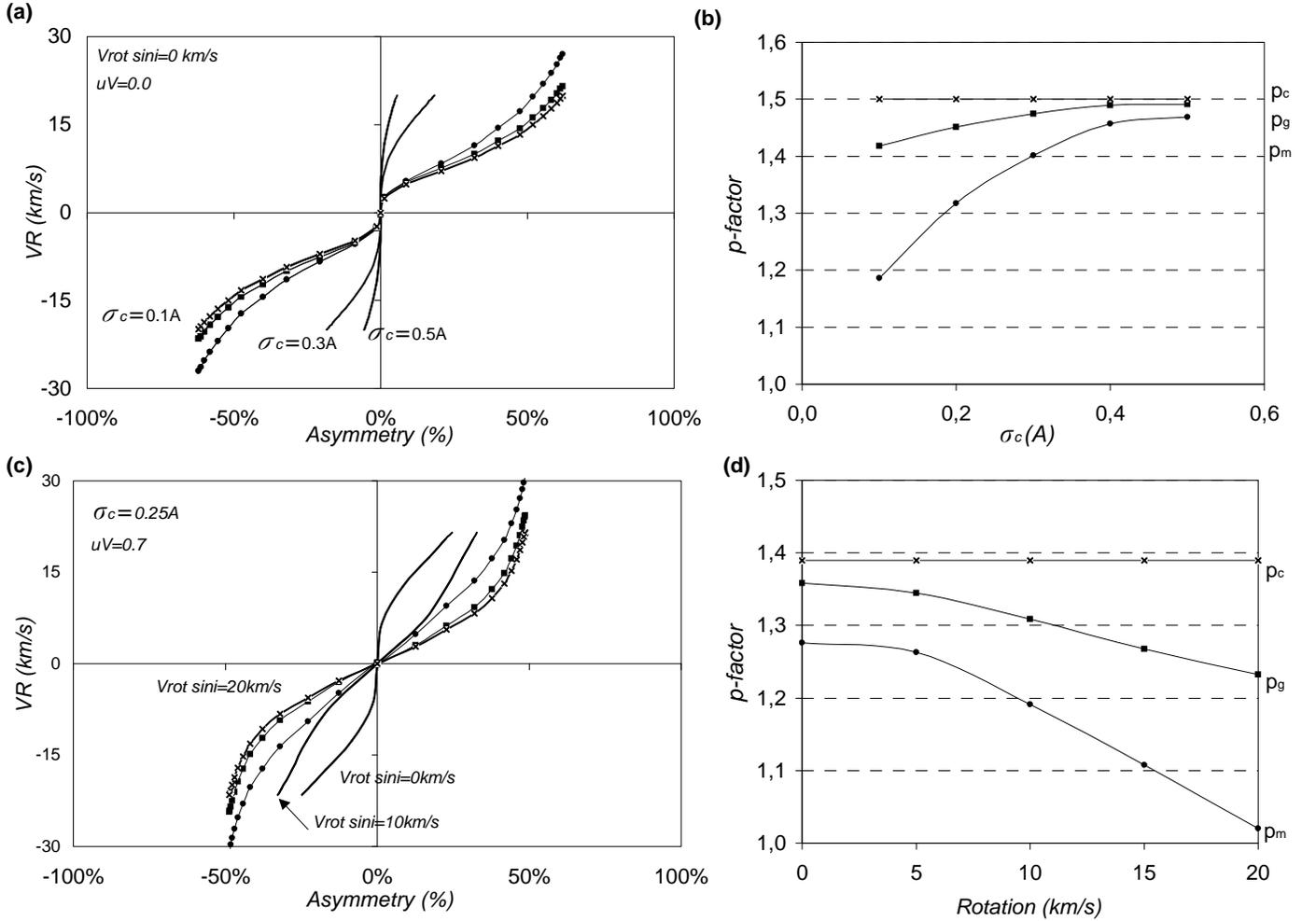}}
\caption{Results of the  geometric model of pulsating star. (a-b)
The radial velocity-asymmetry correlation curves for different $
\sigma_{\mathrm C}$, with no rotation and no limb-darkening (uniform
disk). Points, squares and crosses correspond respectively to the
$RV_{\rm{m}}$, $RV_{\rm{g}}$ and $RV_{\rm{c}}$ radial velocities.
For clarity $RV_{\mathrm g}$ and $RV_{\mathrm m}$ are represented
only for $\sigma_{\mathrm C}=0.1$\,\AA. The solid lines are the
interpolated curves using a cubic spline function. The corresponding
projection factors are represented on diagram (b). (c-d) Same plots
but for different values of the rotation. The {\it $ \sigma_{\mathrm
C}$} and the limb-darkening are respectively of $0.25$ \AA\ and
$0.7$. These RV-A plot are used to interpret HARPS observations.}
\label{Fig_Model}
\end{figure*}

We now consider a pulsation velocity curve defined by:

\begin{equation} \label{Eq_Vpuls}
V_{\rm puls}(\phi_i)=V_{\rm max} \cos(2 \pi \phi_i)
\end{equation}

with a typical value for the maximal pulsation velocity of $V_{\rm
max}=30$ km.s$^{-1}$. This relation which is a poor approximation of
the pulsation velocity curve is only used for the projection factors
determination (see below). It has no incidence on the results (see
Sect \ref{ss_methodology}).  From the synthetic spectral line
profiles, we perform a bi-gaussian fit to derive the four
parameters\, described in Sect. \ref{ss_BIestimator} : $D$,
$\lambda_{\rm{m}}$, $FWHM$ and $A$. Then we derive the
$RV_{\rm{m}}$, $RV_{\rm{g}}$, $RV_{\rm{c}}$ velocities, and the
corresponding radial velocity-asymmetry correlation curves
(hereafter RV-A plot). In Fig. \ref{Fig_Model}, the RV-A plots are
represented for different values of the {\it $ \sigma_{\mathrm C}$}
and rotation parameters. The limb-darkening (considered as constant
with the pulsation phase) has a very small effect in the weighting
of the line profile and thus practically no impact on the RV-A plot.
Applying a classical minimization process between the pulsation and
radial velocities, we have also derived for each set of parameters
the corresponding constant projection factors\,:\, $p_{\mathrm
m}=\frac{V_{\mathrm puls}}{RV_{\rm{m}}}$, $p_{\mathrm
g}=\frac{V_{\mathrm puls}}{RV_{\rm{g}}}$ and $p_{\mathrm
c}=\frac{V_{\mathrm puls}}{RV_{\rm{c}}}$.


\begin{figure}[]
\resizebox{\hsize}{!}{\includegraphics[clip=true]{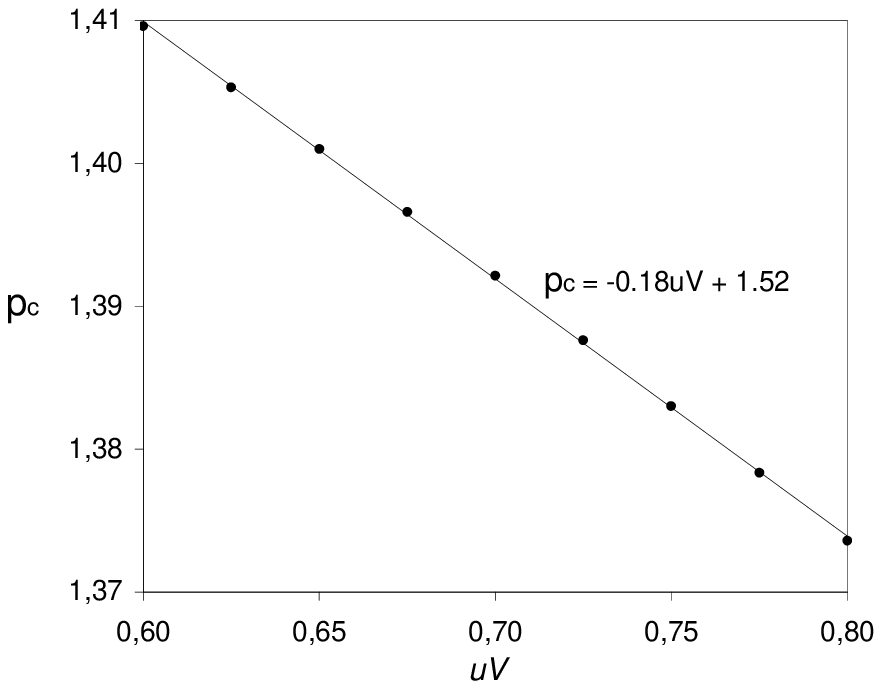}}
\caption{The projection factor corresponding to the centroid
velocity ($p_c$) as a function of the limb-darkening parameter
($u_V$). Dots are the results from the toy model and the solid line
corresponds to the linear approximation ($\chi^2 \simeq 10^{-5}$). }
\label{Fig_pcuv}
\end{figure}

Firstly, we note that the $ \sigma_{\mathrm C}$ of the line and the
rotation have different effects on the slope and/or shape of the
correlation curves.

Secondly, correlation curves are slightly different from one
definition of radial velocity to another. But the interesting point
is that the $RV_{\mathrm{c}}$ velocity does not depend of
$\sigma_{\mathrm C}$ and/or rotation. This behavior is clearly seen
on diagrams \ref{Fig_Model}b and \ref{Fig_Model}d : the centroid
projection factor $p_{\mathrm c}$ is constant with the {\it $
\sigma_{\mathrm C}$} and the rotation while the gaussian and the
minimum projection factors, $p_{\mathrm g}$ and $p_{\mathrm m}$, are
varying. For the Cepheids of our sample the centroid projection
factor ranges from $p_{\mathrm c}=1.40$ ($u_{\mathrm V}=0.64$;
R~TrA) to $p_{\mathrm c}=1.38$ ($u_{\mathrm V}=0.75$; $\ell$~Car),
through the following relation :

\begin{equation} \label{Eq_pc_uv}
p_{\mathrm c}=-0.18 u_{\mathrm V} + 1.52
\end{equation}

This relation is an linear approximation from the geometrical model
(see Fig. \ref{Fig_pcuv}). Note that the geometrical model does not
contain the physics of the pulsations, and thus the relation may not
hold when instead of $u_V$ a more realistic limb-darkening (taking
into account hydrodynamic effects) is used. In particular,
hydrodynamic effects can result in a much larger limb-darkening,
especially at the wavelengths corresponding to spectral line (see
e.g. Marengo et al. 2003).

This behavior is of great importance in the context of the IBW
method. Indeed, the community has often used the $p_{\mathrm
c}=1.36$ value of the projection factor (Burki et al. \cite{b82})
using the gaussian method instead of the centroid method. As seen
here, and already pointed out by Burki et al. \cite{b82}, this
estimator is biased by the rotation velocity, even if Cepheids are
supposed to be slow rotators, and also by the $ \sigma_{\mathrm C}$.
We thus recommend the centroid based methods (spectral observable
and $p$-factor) for the analysis of Cepheid radial velocities. For
the present work, we have therefore chosen the $RV_{\mathrm c}$
definition of the radial velocity. Even though this requires
substantial $S/N$, its advantages outweigh the drawback of spending
more telescope time to acquire the data.

\section{Interpretation}\label{s_interpretation}

\subsection{Methodology} \label{ss_methodology}

\begin{table*}
\begin{center}
\caption[]{ Optimized parameters obtained for each sample Cepheid
through the confrontation of HARPS observations with our geometric
model} \label{Tab_Cepheids_Results}
\begin{tabular}{l|ccccccccc}
\hline \hline \noalign{\smallskip}

stars  &     R TrA     &          S Cru       &        Y Sgr   & $\beta$ Dor       &         $\zeta$ Gem  &   Y Oph        &       RZ Vel        &      l Car        &       RS Pup                          \\
\hline
Period &   3.38925         &   4.68976         &   5.77338         &   9.84262         &   10.14960            &   17.12520            &   20.40020            &   35.551341 & 41.51500 \\
{\it mean} $T_{eff}$$^{\mathrm{a}}$[K] & 6354 & 5995 & 5350 & 5490 & 5727 & 5907 & 5537 & 5091 & 5143 \\
{\it mean}  $\log(g)$$^{\mathrm{a}}$   & 2.0  & 1.9  & 1.0  & 1.8  & 1.5  & 1.5  & 1.5  & 1.5  & 0.4  \\
$u_{\mathrm{V}}^{\mathrm{b}}$ &  0.6371        &       0.6541      &       0.7194        &     0.6999              &  0.6721                 &      0.6514             &  0.6970      &  0.7541      &  0.7121   \\
$ v_{\mathrm{\gamma}}$$^{\mathrm{c}}$ [km.s$^{-1}$]   &  $-13.2$  &  $-7.1$  & $-2.5$ &  $7.4$  & $6.9$    & $-6.6$   & $24.1$  &  $3.6$ & $22.1$ \\
\hline
$ \sigma_{\mathrm C}$~{\tiny (d)} [ \AA\ ]  &     0.29          &   0.27            &        0.27       &       0.23        &     0.23              &  0.20                 &      0.23             &   0.25     & 0.30 \\
$V_{\rm rot} \sin i$~{\tiny (e)} [km.s$^{-1}$] & 15 &  10 &   16 &   6 &    6 &   4 & 3 & 7 & $< 1$ \\
$p_m=\frac{V_{puls}}{RV_m}$  & 1.13  &   1.23 &   1.10  &   1.23 & 1.23 & 1.23  &   1.26 &   1.23 &   1.31 \\
$p_g=\frac{V_{puls}}{RV_g}$  & 1.28  &  1.31  &  1.26 &   1.32 & 1.32 & 1.33  &   1.34 &    1.31 &   1.36 \\
$p_{c}=\frac{V_{puls}}{RV_{c}}$  & 1.40   & 1.40  & 1.39 & 1.39 & 1.40  &  1.40  &  1.39  &  1.38  &  1.39 \\
\hline
$\gamma_{\mathrm O}$~{\tiny (e)}  [$\%$]  & $3.3$ &  $0.7$ &   $ 2.0$ &   $0.2$  &    $-2.4$  &   -  & $-3.2$ &  $ -6.9 $ &  $-6.5 $ \\
$\gamma_{\mathrm C}$~{\tiny (g)} [$\%$] & 3.1       & 4.3  &    0.4  &     2.9 &    0.5 & - &    1.4 &    1.2 &    0.6 \\
$\gamma_{\mathrm O-C}$~{\tiny (h)} [$\%$] & 0.2    & $-3.6$   & 1.6   & $-2.7$  & $-2.9$ & - &  $-4.6$ & $-8.2$ & $-7.1$ \\

\hline \noalign{\smallskip}
\end{tabular}
\end{center}
\begin{list}{}{}

\item[$^{\mathrm{a}}$] $T_{eff}$[K] and $\log(g)$, deduced from
Gieren et al. \cite{gieren98} for R~TrA, S~Scu, Y~Oph and RZ~Vel.
For Y~Sgr, $\beta$~Dor, $\zeta$~Gem, $\ell$~Car, and RS~Pup these
quantities have taken from Cayrel de Strobel et al.
(\cite{Cayrel97}, \cite{Cayrel01}).

\item[$^{\mathrm{b}}$] ${u_\mathrm{V}}$ from Claret et al.
\cite{claret00}

\item[$^{\mathrm{c}}$] ${v_\mathrm{\gamma}}$ from Galactic Cepheid
database (online :
http://www.astro.utoronto.ca/DDO/research/cepheids)

\item[$^{\mathrm{d}}$] Uncertainty on $ \sigma_{\mathrm C}$ is of
about $0.02$\AA.

\item[$^{\mathrm{e}}$] Uncertainty on $V_{\rm rot} \sin i$ is of
about $1$km.s$^{-1}$.

\item[$^{\mathrm{f}}$] $\gamma_O$ [$\%$] is the averaged value of the
observational asymmetry curves. The associated statistical
uncertainties are of the order of $0.3$\%.

\item[$^{\mathrm{g}}$] $\gamma_C$ [$\%$] is the averaged value of the
computed asymmetry curves.

\item[$^{\mathrm{h}}$] $\gamma_{O-C}$ [$\%$] is the average value of the O-C asymmetry curve.

\end{list}

\end{table*}

Modeling results obtained in the previous section are now helpful to
elaborate a strategy in a comparison of observations and models.

Firstly, the effective temperature $T_{\rm eff}$ and the surface
gravity $\log g$ have been used to derive the intensity profile of
stars through linear limb-darkening coefficients $u_{\mathrm{V}}$ of
Claret et al. \cite{claret00} (see Table
\ref{Tab_Cepheids_Results}).

Secondly, we determine the projection factor $p_{\mathrm c}$ using
Eq. \ref{Eq_pc_uv}. The pulsation velocity is then derived through
$V_{\mathrm puls}=p_{\mathrm c} RV_{\mathrm c}$, where $RV_{\mathrm
c}$ is the observational radial velocity corrected from the
heliocentric velocity given in Table 2. The pulsation velocity
$V_{\mathrm puls}$ and the projection factor $p_{\mathrm c}$ (see
Table 2) obtained are not physically realistic, because our model
does not include dynamical effects and in particular velocity
gradients in the atmosphere, nevertheless this procedure imposes the
surimposition of observational and modelled radial velocity curves
$RV_{\mathrm c}$. Moreover, as a very good agreement is observed for
each phase (better than 1\%), it validates the use of a {\it
constant} projection factor ($p_c$). We find also that the poor
description of the pulsation velocity (Eq. \ref{Eq_Vpuls}) used to
derive $p_c$ has no incidence on the resulting modelled $RV_c$
curve. By this procedure, we can thus concentrate only on the
asymmetry, making the interpretation easier. Note that Nardetto et
al. \cite{Nardetto04} already gave an indication of the impact of
velocity gradients on the projection factor, and thus on the
distance determination, in the case of $\delta$ Cep (about $-6\%$).
In Table 2, we also indicate for each star the corresponding
projection factors $p_{\mathrm g} $ and $p_{\mathrm m}$ for
comparison.

Thirdly, $ \sigma_{\mathrm C}$ and $ V_{\mathrm rot} \sin i$ are
determined together from the observational RV-A and FWHM curves. We
first consider the minimum of the observational FWHM curve to obtain
an indication on the value of $\sigma_{\mathrm C}$. We then find the
rotation which gives the best slope and shape for the RV-A curve.
But as the rotation has also an impact on the FWHM (about 0.02\AA),
we have then to slightly readjust $\sigma_{\mathrm C}$ accordingly.
By this process we finally find the best and unique values for
$\sigma_{\mathrm C}$ and $ V_{\mathrm rot} \sin i$.

The uncertainties on $V_{\mathrm rot} \sin i$ and $ \sigma_{\mathrm
C}$, associated to the minimization process, were estimated to be
respectively $1$km.s$^{-1}$ and $0.02$\AA\,. Similar uncertainties
are found if one considers several metallic lines. Note however that
our toy model is too simple to provide secure and precise values of
the rotation, which is the most interesting parameter. In particular
the broadening of the spectral line due to the macro-turbulence can
certainly affect our rotation values (Bersier \& Burki
\cite{Bersier96}). Nevertheless our principal and first objective is
to probe the dynamical effects by a direct comparison of our static
model with observations.

\subsection{Observations Versus Modelisation} \label{ss_Interpretation}

\begin{figure*}
\begin{center}
\includegraphics[width=14cm]{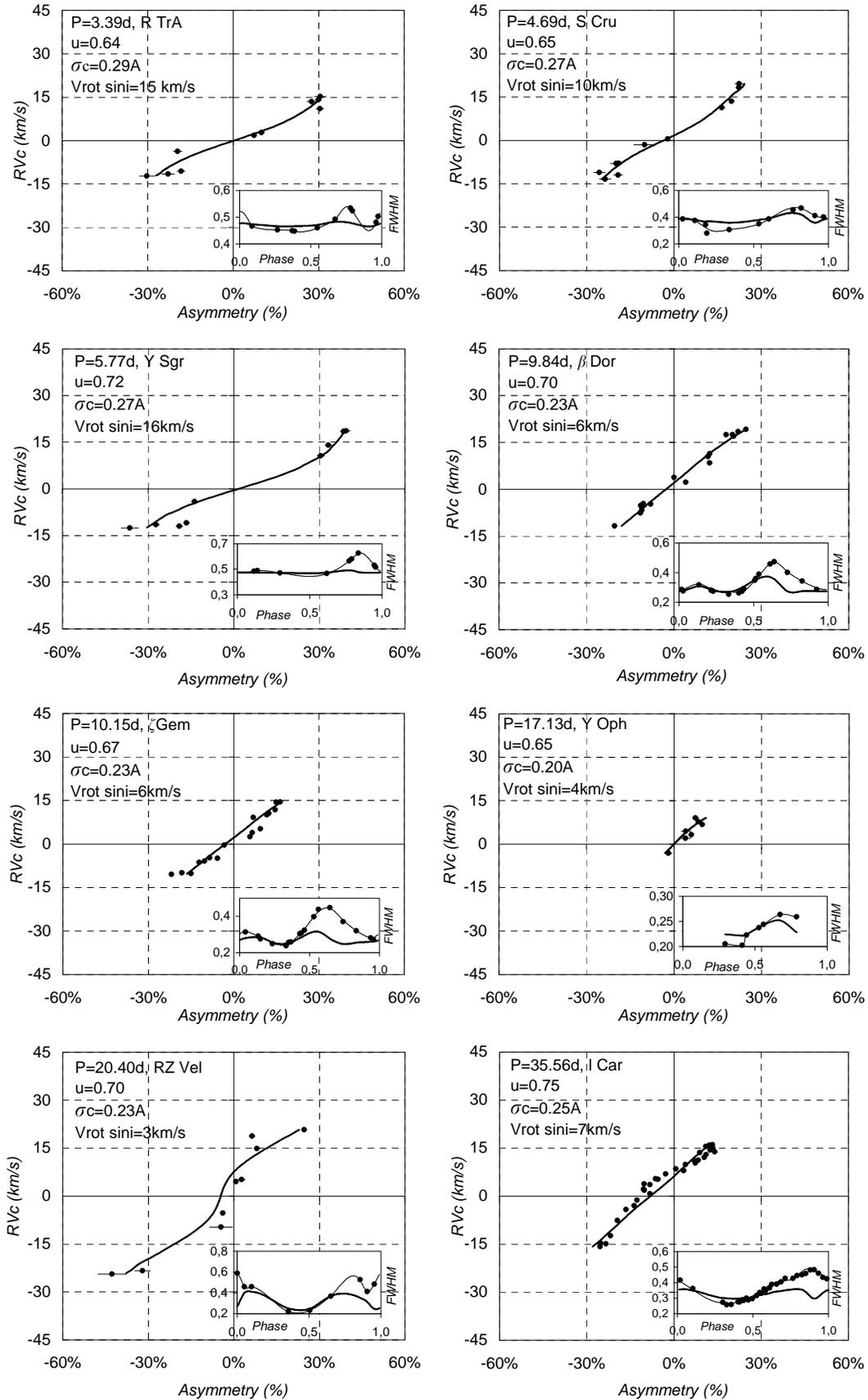}
\end{center}
\caption{Radial velocity ($RV_{\rm c}$) - asymmetry correlation
curves for R~TrA, S~Cru, Y~Sgr, $\beta$~Dor, $\zeta$~Gem, Y~Oph,
RZ~Vel and $\ell$~Car. Dots and bold curves correspond respectively
to observations and models. The statistical uncertainties are
indicated. Note that RV-A plot deduced from the model have been
shifted in asymmetry. The small plot on each diagram correspond to
the comparison of the observational (dots) and model (bold curve)
FWHM.} \label{Fig_cadran}
\end{figure*}

\begin{figure}[h]\resizebox{\hsize}{!}{\includegraphics[clip=true]{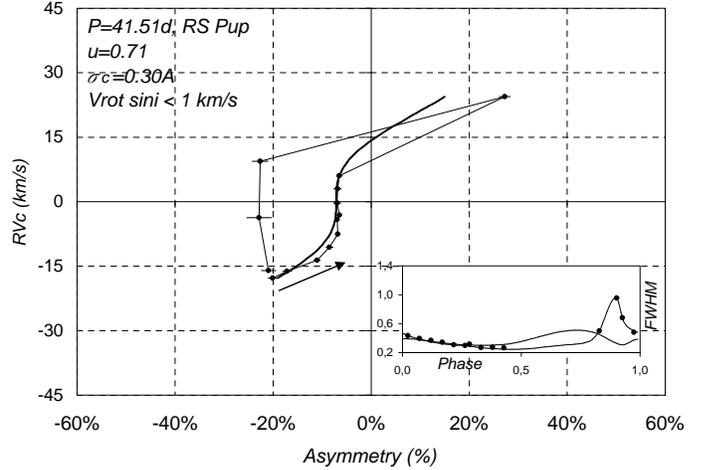}}
\caption{Same as Fig. \ref{Fig_cadran} but for RS~Pup. RS~Pup seems
to be a non-rotating star as requested by the shape of its RV-A
curve. Note also atypical points in observational RV-A plot, which
can certainly be interpreted through the presence of a strong
compression or shock wave in the stellar atmosphere.}
\label{Fig_RSP}
\end{figure}

We now apply our methodology to each Cepheid of our sample. Results
are indicated in Table \ref{Tab_Cepheids_Results}. RV-A plot are
represented on Fig.~\ref{Fig_cadran} and \ref{Fig_RSP}. Note that
RV-A plot deduced from the model have been shifted in asymmetry to
match the observations (this point is discussed in next section).
For R~TrA and Y~Sgr, we can notice a very small slope for the RV-A
plot and a very large value for the observational FWHM. It indicates
a large rotational velocity $V_{\mathrm rot} \sin i$ and a properly
small value for $ \sigma_{\mathrm  C}$ (see Fig. \ref{Fig_Model}ac).
Thus, the corresponding gaussian and minimum projection factors
($p_{\mathrm g}$ and $p_{\mathrm m}$) are lower than for others
stars (see Fig. \ref{Fig_Model}bd). Conversely, for Y~Oph and RZ~Vel
the RV-A plot have relatively large slope while the observational
FWHM is typical (about $0.3$). This has a direct consequence on the
rotation, which is then very small, and on the projection factors
($p_{\mathrm g}$ and $p_{\mathrm m}$) which are then relatively
large. Comparatively, S~Cru, $\beta$ Dor and $\zeta$ Gem can be
considered as intermediate cases. For $\ell$~Car and RS~Pup, we
obtain an atypical RV-A plot which is greatly shifted in asymmetry.
For RS~Pup, we obtain a specific RV-A plot characterized by a strong
curvature which can be interpreted by our geometric model as a very
slow rotation velocity $V_{rot} sin i < 1 $ km.s$^{-1}$. Note that
atypical points which are observed at the top of the RV-A plot are
certainly due to dynamical effects since they corresponds to phases
of outwards acceleration.


\subsection{Discussion}\label{ss_Discussion}

\begin{figure}[]
\resizebox{\hsize}{!}{\includegraphics[clip=true]{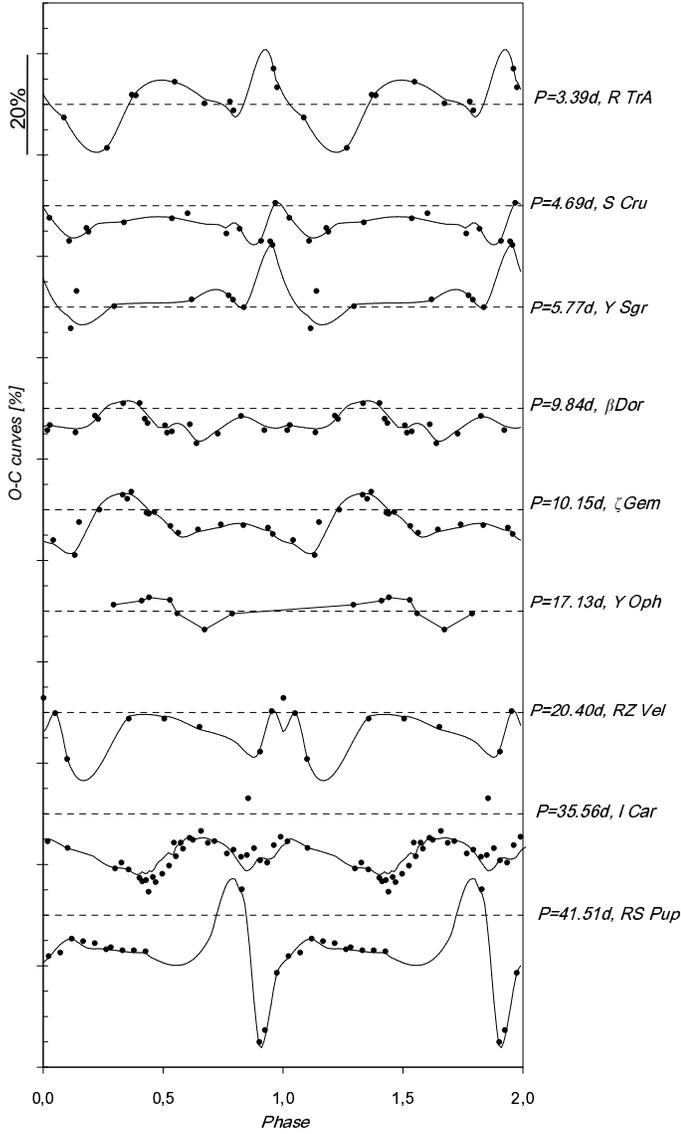}}
\caption{Difference of the Observational and Computed asymmetry
curves (O-C curves) for each stars. Curves are arbitrarily shifted.
The horizontal dotted lines corresponds to a zero asymmetry for each
star.} \label{Fig_discussion1}
\end{figure}

As observed in the particular case of $\ell$~Car and RS~Pup, an
important systematic shift in asymmetry can be present between
observations and models. We define respectively $\gamma_O$ and
$\gamma_C$ the averaged value of the observational and computed
asymmetry curves [in $\%$]. Note that the phases are sampled in the
same way for data and model. Results are indicated in
Table\,\ref{Tab_Cepheids_Results}. We have also calculated for each
star the residuals between the observational and computed asymmetry
curves, noted O-C curves (Fig. \ref{Fig_discussion1}). We define
$\gamma_{\mathrm O-C}$, the average value of these residual curves.
These O-C asymmetry curves contain the whole dynamical information
present in the observational asymmetry, mainly\,:\, the
limb-darkening variation in the spectral line and with the pulsation
phase, the micro- and macro- turbulence, velocity gradient and
temperature effects. For R~TrA, S~Cru, Y~Sgr, RZ~Vel and RS~Pup, we
note a bump in the O-C asymmetry curves which is approximately
linked to the cross of the compression wave just after the maximum
contraction velocity (see Fig. \ref{Fig_RVmShifted}). However
$\beta$~Dor, $\zeta$~Gem and $\ell$~Car do not present such bump,
which may be interpreted as the presence of a very small compression
wave. In the case of Y~Oph the phase sampling seems insufficient to
conclude. Consistent hydrodynamical model would be helpful to
confirm these results.

\begin{figure}[]
\resizebox{\hsize}{!}{\includegraphics[clip=true]{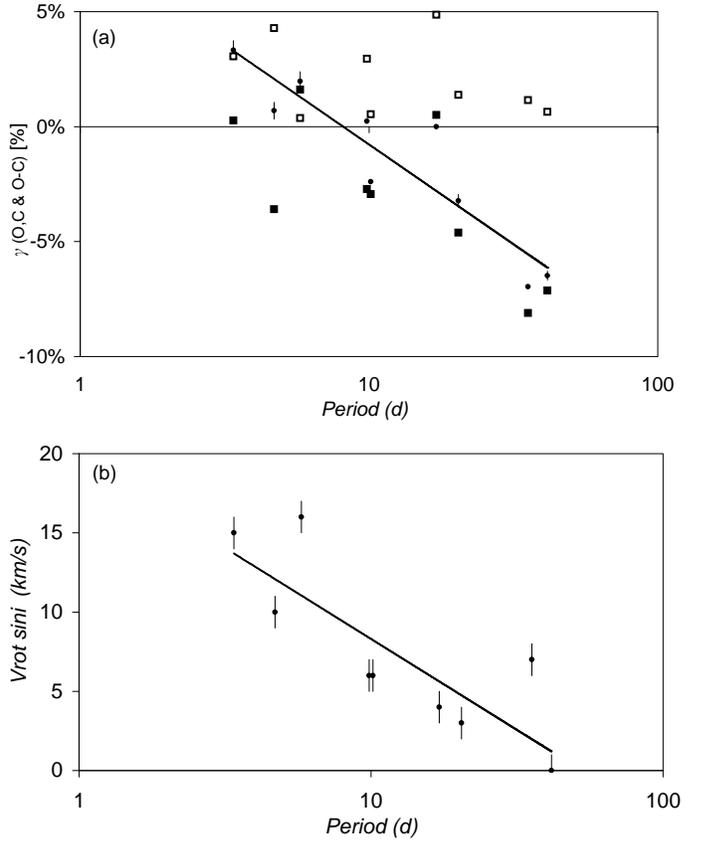}}
\caption{(a) Average values of the observational (black circles) and
computed (open squares) asymmetry curves, together with the
  $\gamma_{O-C}$ (filled squares) average values as a function of the
pulsation period. (b) Dependence of the projected rotation velocity
with the pulsation period. } \label{Fig_discussion2}
\end{figure}

$\gamma_{O}$, $\gamma_{C}$ and $\gamma_{O-C}$ are represented as a
function of the pulsation period on Fig. \ref{Fig_discussion2}a. The
open squares represent $\gamma_{C}$. We want to emphasize here that
our model produces asymmetry curves with {\it non}-zero average
value. Indeed, it is a natural consequence of the shape of the
observational radial velocity curve used to derive the pulsation
velocity. We find a similar behavior for all stars independently of
the period.

The shifts obtained on the observational asymmetry curves
($\gamma_{O}$) show a very interesting linear dependence with the
logarithm of the pulsation period:

\begin{equation} \label{Eq_Shift}
\gamma_{O}=(-10.7 \pm 0.1) \log(P) + (9.7 \pm 0.2) \mbox{ [in \%] }
\end{equation}

Moreover we note that the dependence of $\gamma_{O-C}$ with the
pulsation period is very similar to the one of $\gamma_{O}$. We can
conclude that this behavior is related to the dynamical effects in
the atmosphere, which are not taken into account in our toy model.
This can be explained by the fact that long-period Cepheids have
extended atmosphere and consequently strong velocity gradient (see
for example the case of RS~Pup mentioned above). Thus, the line
forming region can be seriously perturbed leading to a systematic
shift in asymmetry (Albrow \& Cottrell \cite{Albrow94}). However,
such an interpretation remains tricky and needs confirmation.
Forthcoming hydrodynamical models are likely to bring out important
insight in this field.

From results of Table\,\ref{Tab_Cepheids_Results}, it appears also
that the projected rotational velocity varies as a function of the
pulsation period (Fig.~\ref{Fig_discussion2}b). We obtain the
following relationship:

\begin{equation} \label{Eq_Rotation}
V_{\mathrm rot} \sin i = (-11.5 \pm 0.9) \log(P) + (19.8 \pm 1.0)
\mbox{ [in km.s$^{-1}$]}
\end{equation}

The projected rotation is an important parameter which can be used,
for example, to study evolution of Cepheids together with their mass
loss. However, note again that our toy model does not include the
physics of the pulsations and it is also very difficult to separate
the rotation and macroturbulence effects in the resulting broadening
of the spectral line. Thus this relation has to be considered very
carefully as it is certainly model dependent.

\section{Conclusion}\label{s_Conclusion}

We have presented HARPS high spectral resolution (R=120000)
observations of nine galactic Cepheids having a good period sampling
($P=3.39$d to $P=41.52$d). We fit spectral line profile with an
asymmetric bi-gaussian to derive radial velocity, $FWHM$ and line
asymmetry for all stars. The presence of a very important
compression or shock wave in the case of RS Pup, the longest period
Cepheid of our sample has been identified. We have also translated
the measured spectroscopic quantities into meaningful correlation
curves between radial velocity and asymmetry.

A simple geometric model providing synthetic spectral lines,
including limb-darkening, the $ \sigma_{\mathrm C}$ and the
projected rotation velocity is then used to interpret these
correlations curves.

Firstly, we find that the centroid projection factor ($p_{\mathrm
c}$) is independent of $\sigma_{\mathrm C}$ and the rotation
velocity. This projection factor is thus certainly the best one to
use in the context of the Baade-Wesselink method.

Secondly, we find for each stars an optimized set of parameters
which allows to reproduce observational radial velocity - asymmetry
correlation curves. In particular, we find a dependence of the
derived projected rotation velocities with the period of the star :
$V_{\mathrm rot} \sin i= (-11.5 \pm 0.9) \log(P) + (19.8 \pm 1.0)
\mbox{ [in km.s$^{-1}$]}$.

Finally, by comparing the outputs of our static models and the
observed quantities, we gain access to dynamical effects. In
particular, we found that long-period Cepheids with strong velocity
gradient, like RS~Pup, have a systematic shift in their asymmetry
curve. We thus derived a linear relation between the observational
shift in asymmetry and the logarithm of the period :
$\gamma_{O}=(-10.7 \pm 0.1) \log(P) + (9.7 \pm 0.2) \mbox{ [in \%]
}$. A detailed interpretation of these empirical relation is very
difficult, but forthcoming hydrodynamical models are likely to bring
out important insight in this field.

In conclusion, line asymmetry, which contains most of the physics
involved in Cepheid atmosphere, is an important tool. But additional
hydrodynamical considerations together with a multi-lines study are
now required to have a better understanding of the dynamical
processes present in Cepheid atmosphere and in particular to
determine realistic projection factors including velocity gradients.

\begin{acknowledgements}
Based on observations collected at La Silla observatory, Chile, in
the framework of European Southern Observatory's programs 072.D-0419
and 073.D-0136. This research has made use of the SIMBAD and VIZIER
databases at CDS, Strasbourg (France). We thanks David Chapeau for
his helpful collaboration concerning computing aspects, Olivier
Chesneau and Philippe Stee for their careful reading of the
manuscript, as well as Vincent Coud\'e du Foresto and Andrei Fokin
for useful discussions.

\end{acknowledgements}

\begin{table*}
\begin{center}
\caption[]{HARPS observations results for R~TrA, S~Cru and Y~Sgr.
\label{Tab_cepheids_results1}}
\begin{tabular}{cccccccccccc}
\hline \hline \noalign{\smallskip}

  JD$_{\mathrm c}$  &  phase   &   Cy.     &   Sp.         &  $RV_{\mathrm g}$  &   $RV_{\mathrm m}$   &   $ RV_{\mathrm c}$       &   $FWHM$   & $D$  & $A$ & $SNR$ & $\chi_{\mathrm red}^{2}$  \\
  {\tiny (a)}  &  {\tiny (b)}   &   {\tiny (c)}      &   {\tiny (d)}          &  {\tiny (e)}   &   {\tiny (f)}   &   {\tiny (g)}       &   {\tiny (h)}   & {\tiny (i)}  & {\tiny (j)} & {\tiny (k)} & {\tiny (l)}  \\

\hline \noalign{\smallskip}
\multicolumn{12}{c}{R~TrA}\\
\noalign{\smallskip}

202.53  &   0.09    &   14  &   1   &   -26.74  $\pm$   0.07    &   -29.88  $\pm$   0.17    &   -25.43  $\pm$   1.16    &   0.467   $\pm$   0.009   &   0.14    &   -30.3  $\pm$   2.7    &   231 &   2.0   \\
206.53  &   0.27    &   15  &   1   &   -17.70  $\pm$   0.05    &   -19.71  $\pm$   0.13    &   -16.92  $\pm$   0.91    &   0.452   $\pm$   0.005   &   0.17    &   -19.5  $\pm$   1.5    &   224 &   1.6   \\
152.65  &   0.37    &   1   &   2   &   -11.18  $\pm$   0.03    &   -10.41  $\pm$   0.08    &   -11.39  $\pm$   0.56    &   0.449   $\pm$   0.002   &   0.19    &   7.4    $\pm$   0.8    &   241 &   1.6   \\
203.55  &   0.39    &   14  &   1   &   -10.08  $\pm$   0.04    &   -9.06   $\pm$   0.11    &   -10.47  $\pm$   0.71    &   0.447   $\pm$   0.003   &   0.18    &   9.9    $\pm$   1.1    &   257 &   1.1   \\
156.65  &   0.55    &   2   &   2   &   -0.98   $\pm$   0.04    &   2.20    $\pm$   0.09    &   -2.15   $\pm$   0.52    &   0.460   $\pm$   0.005   &   0.19    &   30.4   $\pm$   1.3    &   209 &   1.0   \\
204.52  &   0.67    &   14  &   1   &   3.29    $\pm$   0.06    &   6.70    $\pm$   0.15    &   2.02    $\pm$   0.74    &   0.492   $\pm$   0.008   &   0.17    &   30.6   $\pm$   2.0    &   207 &   1.4   \\
150.65  &   0.78    &   1   &   2   &   2.26    $\pm$   0.04    &   5.89    $\pm$   0.09    &   0.89    $\pm$   0.50    &   0.534   $\pm$   0.005   &   0.16    &   29.9   $\pm$   1.1    &   255 &   1.3   \\
201.54  &   0.79    &   14  &   1   &   1.34    $\pm$   0.06    &   4.64    $\pm$   0.14    &   0.30    $\pm$   0.74    &   0.524   $\pm$   0.007   &   0.16    &   27.5   $\pm$   1.6    &   247 &   1.7   \\
154.65  &   0.96    &   2   &   2   &   -24.46  $\pm$   0.05    &   -26.46  $\pm$   0.13    &   -23.76  $\pm$   0.79    &   0.481   $\pm$   0.004   &   0.12    &   -18.3  $\pm$   1.4    &   241 &   1.2   \\
205.54  &   0.98    &   15  &   1   &   -25.60  $\pm$   0.08    &   -28.18  $\pm$   0.19    &   -24.71  $\pm$   1.24    &   0.503   $\pm$   0.007   &   0.12    &   -22.8  $\pm$   2.1    &   243 &   1.4   \\

\noalign{\smallskip}
\multicolumn{12}{c}{S~Cru}\\
\noalign{\smallskip}

207.46  &   0.03    &   3   &   1   &   -21.58  $\pm$   0.05    &   -23.61  $\pm$   0.13    &   -20.47  $\pm$   1.01    &   0.386   $\pm$   0.005   &   0.16    &   -23.6  $\pm$   2.0    &   230 &   1.5   \\
151.56  &   0.11    &   1   &   1   &   -19.32  $\pm$   0.05    &   -21.43  $\pm$   0.12    &   -18.20  $\pm$   0.93    &   0.375   $\pm$   0.005   &   0.18    &   -25.6  $\pm$   2.0    &   214 &   1.4   \\
203.49  &   0.18    &   3   &   2   &   -15.82  $\pm$   0.03    &   -17.26  $\pm$   0.07    &   -14.98  $\pm$   0.50    &   0.344   $\pm$   0.002   &   0.21    &   -19.0  $\pm$   1.0    &   224 &   1.7   \\
156.63  &   0.19    &   1   &   1   &   -15.65  $\pm$   0.04    &   -16.90  $\pm$   0.10    &   -15.07  $\pm$   0.52    &   0.281   $\pm$   0.004   &   0.19    &   -19.8  $\pm$   1.9    &   221 &   2.4   \\
152.63  &   0.34    &   1   &   1   &   -6.94   $\pm$   0.02    &   -7.08   $\pm$   0.06    &   -6.54   $\pm$   0.46    &   0.306   $\pm$   0.001   &   0.26    &   -2.1   $\pm$   0.8    &   255 &   1.9   \\
153.57  &   0.54    &   1   &   1   &   4.61    $\pm$   0.03    &   5.87    $\pm$   0.08    &   4.21    $\pm$   0.62    &   0.351   $\pm$   0.002   &   0.26    &   16.5   $\pm$   1.1    &   209 &   2.5   \\
205.47  &   0.60    &   3   &   1   &   7.09    $\pm$   0.03    &   8.77    $\pm$   0.06    &   6.44    $\pm$   0.59    &   0.386   $\pm$   0.002   &   0.26    &   19.9   $\pm$   0.9    &   269 &   2.5   \\
154.64  &   0.76    &   1   &   1   &   13.40   $\pm$   0.05    &   15.66   $\pm$   0.12    &   12.48   $\pm$   1.06    &   0.454   $\pm$   0.005   &   0.23    &   22.5   $\pm$   1.4    &   181 &   1.6   \\
206.48  &   0.82    &   3   &   1   &   12.07   $\pm$   0.04    &   14.42   $\pm$   0.09    &   11.17   $\pm$   0.59    &   0.469   $\pm$   0.004   &   0.20    &   22.4   $\pm$   1.0    &   285 &   1.5   \\
150.63  &   0.91    &   1   &   1   &   -9.32   $\pm$   0.14    &   -10.19  $\pm$   0.35    &   -8.58   $\pm$   1.37    &   0.412   $\pm$   0.008   &   0.17    &   -10.1  $\pm$   3.7    &   87  &   1.4   \\
202.49  &   0.97    &   3   &   1   &   -19.91  $\pm$   0.05    &   -21.62  $\pm$   0.12    &   -19.05  $\pm$   0.70    &   0.401   $\pm$   0.004   &   0.15    &   -19.2  $\pm$   1.5    &   287 &   2.0   \\

\noalign{\smallskip}
\multicolumn{12}{c}{Y~Sgr}\\
\noalign{\smallskip}

204.63  &   0.12    &   10  &   2   &   -16.53  $\pm$   0.06    &   -20.47  $\pm$   0.15    &   -15.07  $\pm$   0.86    &   0.485   $\pm$   0.011   &   0.15    &   -36.4  $\pm$   3.0    &   160 &   1.1   \\
152.80  &   0.14    &   1   &   2   &   -15.07  $\pm$   0.04    &   -18.08  $\pm$   0.09    &   -14.02  $\pm$   0.56    &   0.488   $\pm$   0.004   &   0.16    &   -27.2  $\pm$   1.3    &   251 &   1.4   \\
205.67  &   0.30    &   10  &   1   &   -6.93   $\pm$   0.05    &   -8.45   $\pm$   0.12    &   -6.56   $\pm$   0.65    &   0.472   $\pm$   0.003   &   0.19    &   -13.7  $\pm$   1.1    &   244 &   2.6   \\
149.80  &   0.62    &   1   &   2   &   9.37    $\pm$   0.04    &   12.58   $\pm$   0.09    &   8.13    $\pm$   0.62    &   0.467   $\pm$   0.005   &   0.21    &   30.5   $\pm$   1.3    &   178 &   1.0   \\
202.65  &   0.77    &   10  &   2   &   18.50   $\pm$   0.04    &   23.46   $\pm$   0.09    &   16.08   $\pm$   0.71    &   0.565   $\pm$   0.007   &   0.19    &   39.5   $\pm$   1.4    &   231 &   1.5   \\
150.79  &   0.79    &   1   &   2   &   18.31   $\pm$   0.03    &   23.31   $\pm$   0.08    &   15.98   $\pm$   0.61    &   0.581   $\pm$   0.006   &   0.18    &   38.5   $\pm$   1.1    &   270 &   1.8   \\
156.83  &   0.84    &   2   &   2   &   13.31   $\pm$   0.04    &   18.07   $\pm$   0.11    &   11.56   $\pm$   0.63    &   0.626   $\pm$   0.007   &   0.16    &   33.2   $\pm$   1.2    &   255 &   1.6   \\
203.65  &   0.95    &   10  &   2   &   -14.27  $\pm$   0.04    &   -16.27  $\pm$   0.11    &   -13.47  $\pm$   0.50    &   0.530   $\pm$   0.003   &   0.13    &   -16.5  $\pm$   1.0    &   288 &   1.4   \\
151.75  &   0.96    &   1   &   2   &   -15.22  $\pm$   0.05    &   -17.50  $\pm$   0.12    &   -14.51  $\pm$   0.61    &   0.517   $\pm$   0.004   &   0.13    &   -19.1  $\pm$   1.2    &   254 &   1.1   \\

 \hline \hline \noalign{\smallskip}
\end{tabular}
\end{center}

\begin{list}{}{}
\item {\tiny (a)} JD$_{\mathrm c}$, average Julian date of
observation defined by $JD_{\mathrm c}=JD-2453000$ [in days].

\item {\tiny (b)} phase,  averaged pulsation phase of
observation. For ephemeris see Table \ref{Tab_Cepheids_Parameters}.

\item {\tiny (c)} Cy., pulsating cycle of the star
corresponding to observation.

\item {\tiny (d)} Sp., number of spectra associated to
observation. Results corresponding to these spectra are averaged.

\item {\tiny (e)} $RV_{\mathrm g}$, gaussian fit radial velocity
and the associated error barre [in $km.s^{-1}$].

\item {\tiny (f)} $RV_{\mathrm m}$,  minimum radial velocity
derived from the bi-gaussian fit [in $km.s^{-1}$].

\item {\tiny (g)} $ RV_{\mathrm c}$, radial velocity
corresponding to the first moment of the spectral line [in
$km.s^{-1}$].

\item {\tiny (h)} $FWHM$, Full-Width at Half-Maximum
derived from the bi-gaussian fit [in Angstroms].

\item {\tiny (i)} $D$, line depth derived from the
bi-gaussian fit [no dimension]. Errors bars are not indicated but of
the order of $10^{-4}$.

\item {\tiny (j)} $A$, asymmetry derived from the
bi-gaussian fit [in percentage].

\item {\tiny (k)} $SNR$, observational spectral line signal to
noise ratio.

\item {\tiny (l)} $\chi_{\mathrm red}^{2}$, reduced  $\chi^{2}$
factor corresponding to the bi-gaussian fit.

\end{list}
\end{table*}

\begin{table*}
\begin{center}
\caption[]{HARPS observations results for $\beta$~Dor,
$\zeta$~Gem, Y~Oph, and RZ~Vel. See
Table\ref{Tab_cepheids_results1} for legend.
\label{Tab_cepheids_results2}}
\begin{tabular}{cccccccccccc}
\hline \hline \noalign{\smallskip}

  JD$_{\mathrm c}$  &  phase   &   Cy.     &   Sp.         &  $RV_{\mathrm g}$  &   $RV_{\mathrm m}$   &   $ RV_{\mathrm c}$       &   $FWHM$   & $D$  & $A$ & $SNR$ & $\chi_{\mathrm red}^{2}$  \\
  {\tiny (a)}  &  {\tiny (b)}   &   {\tiny (c)}      &   {\tiny (d)}          &  {\tiny (e)}   &   {\tiny (f)}   &   {\tiny (g)}       &   {\tiny (h)}   & {\tiny (i)}  & {\tiny (j)} & {\tiny (k)} & {\tiny (l)}  \\

\hline

\noalign{\smallskip}
\multicolumn{12}{c}{$\beta$~Dor}\\
\noalign{\smallskip}

21.68   &   0.02    &   1   &   4   &   1.70    $\pm$   0.01    &   0.99    $\pm$   0.02    &   2.17    $\pm$   0.14    &   0.286   $\pm$   0.001   &   0.23    &   -11.3  $\pm$   0.4    &   345 &   3.7   \\
31.64   &   0.03    &   2   &   3   &   1.35    $\pm$   0.01    &   0.68    $\pm$   0.02    &   1.64    $\pm$   0.12    &   0.275   $\pm$   0.001   &   0.23    &   -11.0  $\pm$   0.4    &   404 &   2.4   \\
32.68   &   0.14    &   2   &   3   &   -5.16   $\pm$   0.01    &   -6.59   $\pm$   0.03    &   -4.33   $\pm$   0.19    &   0.318   $\pm$   0.001   &   0.24    &   -20.3  $\pm$   0.6    &   298 &   2.1   \\
23.64   &   0.22    &   1   &   4   &   -0.73   $\pm$   0.01    &   -1.42   $\pm$   0.01    &   -0.19   $\pm$   0.11    &   0.280   $\pm$   0.001   &   0.30    &   -11.4  $\pm$   0.2    &   423 &   4.9   \\
33.61   &   0.23    &   2   &   3   &   0.15    $\pm$   0.01    &   -0.49   $\pm$   0.02    &   0.71    $\pm$   0.12    &   0.275   $\pm$   0.001   &   0.31    &   -11.0  $\pm$   0.3    &   443 &   7.9   \\
34.64   &   0.33    &   2   &   2   &   9.67    $\pm$   0.01    &   9.90    $\pm$   0.02    &   9.68    $\pm$   0.22    &   0.253   $\pm$   0.001   &   0.36    &   4.1    $\pm$   0.3    &   330 &   2.5   \\
15.62   &   0.40    &   1   &   3   &   16.16   $\pm$   0.01    &   16.85   $\pm$   0.02    &   15.85   $\pm$   0.28    &   0.261   $\pm$   0.001   &   0.35    &   12.3   $\pm$   0.4    &   262 &   2.4   \\
25.68   &   0.42    &   2   &   3   &   18.10   $\pm$   0.01    &   18.80   $\pm$   0.02    &   17.86   $\pm$   0.24    &   0.273   $\pm$   0.001   &   0.33    &   11.8   $\pm$   0.3    &   399 &   3.3   \\
35.64   &   0.44    &   3   &   2   &   19.09   $\pm$   0.01    &   19.87   $\pm$   0.02    &   18.80   $\pm$   0.35    &   0.290   $\pm$   0.001   &   0.33    &   12.3   $\pm$   0.4    &   337 &   2.0   \\
16.67   &   0.51    &   1   &   3   &   24.95   $\pm$   0.01    &   26.53   $\pm$   0.02    &   24.41   $\pm$   0.38    &   0.347   $\pm$   0.001   &   0.28    &   20.5   $\pm$   0.4    &   352 &   3.2   \\
26.59   &   0.52    &   2   &   2   &   25.48   $\pm$   0.01    &   27.09   $\pm$   0.02    &   24.91   $\pm$   0.34    &   0.359   $\pm$   0.001   &   0.27    &   20.2   $\pm$   0.3    &   473 &   4.9   \\
36.64   &   0.54    &   3   &   2   &   26.61   $\pm$   0.01    &   28.52   $\pm$   0.04    &   25.83   $\pm$   0.54    &   0.388   $\pm$   0.001   &   0.26    &   22.1   $\pm$   0.5    &   336 &   2.2   \\
17.69   &   0.61    &   1   &   3   &   27.57   $\pm$   0.02    &   30.11   $\pm$   0.04    &   26.60   $\pm$   0.51    &   0.457   $\pm$   0.002   &   0.22    &   24.8   $\pm$   0.5    &   303 &   2.2   \\
37.64   &   0.64    &   3   &   2   &   25.54   $\pm$   0.02    &   27.46   $\pm$   0.04    &   24.86   $\pm$   0.45    &   0.473   $\pm$   0.001   &   0.20    &   18.0   $\pm$   0.5    &   409 &   3.1   \\
28.67   &   0.73    &   2   &   3   &   11.09   $\pm$   0.01    &   11.09   $\pm$   0.03    &   11.15   $\pm$   0.20    &   0.401   $\pm$   0.001   &   0.21    &   0.1    $\pm$   0.3    &   456 &   2.7   \\
29.63   &   0.83    &   2   &   4   &   2.24    $\pm$   0.01    &   1.65    $\pm$   0.02    &   2.60    $\pm$   0.12    &   0.343   $\pm$   0.001   &   0.23    &   -7.9   $\pm$   0.2    &   472 &   6.5   \\
30.59   &   0.92    &   2   &   3   &   2.36    $\pm$   0.01    &   1.71    $\pm$   0.02    &   2.77    $\pm$   0.12    &   0.286   $\pm$   0.001   &   0.24    &   -10.5  $\pm$   0.3    &   455 &   5.0   \\

\noalign{\smallskip}
\multicolumn{12}{c}{$\zeta$~Gem}\\
\noalign{\smallskip}

32.70   &   0.04    &   2   &   3   &   -3.82   $\pm$   0.02    &   -5.04   $\pm$   0.04    &   -3.05   $\pm$   0.33    &   0.313   $\pm$   0.001   &   0.27    &   -18.2   $\pm$   0.7     &   196 &   2.9   \\
33.62   &   0.14    &   2   &   3   &   -4.67   $\pm$   0.01    &   -6.02   $\pm$   0.02    &   -3.55   $\pm$   0.19    &   0.292   $\pm$   0.001   &   0.31    &   -21.9   $\pm$   0.4     &   330 &   7.7   \\
23.65   &   0.15    &   1   &   4   &   -4.00   $\pm$   0.01    &   -4.89   $\pm$   0.02    &   -3.35   $\pm$   0.13    &   0.276   $\pm$   0.001   &   0.31    &   -15.0   $\pm$   0.3     &   338 &   4.2   \\
34.65   &   0.23    &   2   &   3   &   1.53    $\pm$   0.01    &   1.23    $\pm$   0.02    &   1.97    $\pm$   0.15    &   0.248   $\pm$   0.001   &   0.37    &   -5.7    $\pm$   0.3     &   334 &   6.0   \\
35.65   &   0.34    &   2   &   3   &   9.44    $\pm$   0.01    &   9.73    $\pm$   0.02    &   9.41    $\pm$   0.18    &   0.239   $\pm$   0.001   &   0.39    &   5.8     $\pm$   0.3     &   299 &   2.6   \\
25.69   &   0.35    &   1   &   3   &   10.86   $\pm$   0.01    &   11.22   $\pm$   0.03    &   10.84   $\pm$   0.35    &   0.256   $\pm$   0.001   &   0.37    &   6.6     $\pm$   0.4     &   195 &   1.9   \\
15.71   &   0.37    &   1   &   3   &   12.29   $\pm$   0.01    &   12.81   $\pm$   0.02    &   12.12   $\pm$   0.25    &   0.259   $\pm$   0.001   &   0.38    &   9.4     $\pm$   0.3     &   253 &   1.6   \\
36.66   &   0.43    &   2   &   2   &   17.10   $\pm$   0.01    &   17.87   $\pm$   0.03    &   16.92   $\pm$   0.46    &   0.304   $\pm$   0.001   &   0.34    &   11.7    $\pm$   0.5     &   255 &   2.3   \\
26.60   &   0.44    &   2   &   3   &   17.69   $\pm$   0.01    &   18.53   $\pm$   0.02    &   17.48   $\pm$   0.29    &   0.308   $\pm$   0.001   &   0.33    &   12.4    $\pm$   0.3     &   353 &   3.1   \\
16.69   &   0.46    &   1   &   3   &   19.06   $\pm$   0.02    &   20.11   $\pm$   0.04    &   18.71   $\pm$   0.50    &   0.324   $\pm$   0.001   &   0.31    &   14.6    $\pm$   0.6     &   189 &   1.4   \\
37.66   &   0.53    &   3   &   2   &   21.81   $\pm$   0.01    &   23.27   $\pm$   0.03    &   21.40   $\pm$   0.45    &   0.397   $\pm$   0.001   &   0.28    &   16.4    $\pm$   0.4     &   341 &   3.1   \\
17.70   &   0.56    &   1   &   3   &   21.79   $\pm$   0.01    &   23.25   $\pm$   0.02    &   21.35   $\pm$   0.33    &   0.439   $\pm$   0.001   &   0.27    &   14.9    $\pm$   0.2     &   446 &   3.1   \\
28.68   &   0.62    &   2   &   2   &   16.11   $\pm$   0.02    &   16.81   $\pm$   0.06    &   16.02   $\pm$   0.59    &   0.449   $\pm$   0.001   &   0.24    &   7.0     $\pm$   0.6     &   243 &   1.4   \\
29.64   &   0.74    &   2   &   2   &   6.25    $\pm$   0.03    &   5.98    $\pm$   0.07    &   6.53    $\pm$   0.53    &   0.372   $\pm$   0.001   &   0.26    &   -3.3    $\pm$   0.8     &   169 &   1.7   \\
30.60   &   0.84    &   2   &   3   &   1.64    $\pm$   0.01    &   1.06    $\pm$   0.02    &   2.09    $\pm$   0.16    &   0.321   $\pm$   0.001   &   0.28    &   -8.5    $\pm$   0.3     &   407 &   7.2   \\
31.64   &   0.94    &   2   &   3   &   0.49    $\pm$   0.01    &   -0.13   $\pm$   0.02    &   0.98    $\pm$   0.15    &   0.282   $\pm$   0.001   &   0.28    &   -10.2   $\pm$   0.3     &   372 &   8.3   \\
21.70   &   0.96    &   1   &   5   &   0.18    $\pm$   0.01    &   -0.54   $\pm$   0.03    &   0.59    $\pm$   0.15    &   0.271   $\pm$   0.001   &   0.27    &   -12.1   $\pm$   0.5     &   229 &   2.3   \\

\noalign{\smallskip}
\multicolumn{12}{c}{ Y~Oph}\\
\noalign{\smallskip}

216.75  &   0.29    &   4   &   1   &   -9.75   $\pm$   0.02    &   -9.83   $\pm$   0.05    &   -9.72   $\pm$   0.46    &   0.205   $\pm$   0.001   &   0.32    &   -1.9    $\pm$   1.1     &   189 &   1.2   \\
201.63  &   0.41    &   4   &   1   &   -4.48   $\pm$   0.03    &   -4.31   $\pm$   0.08    &   -4.57   $\pm$   0.43    &   0.202   $\pm$   0.002   &   0.31    &   4.0     $\pm$   1.8     &   119 &   1.4   \\
150.78  &   0.44    &   1   &   1   &   -3.15   $\pm$   0.02    &   -2.87   $\pm$   0.04    &   -3.37   $\pm$   0.27    &   0.223   $\pm$   0.001   &   0.33    &   5.9     $\pm$   0.7     &   262 &   2.9   \\
203.65  &   0.53    &   4   &   1   &   0.57    $\pm$   0.02    &   1.06    $\pm$   0.04    &   0.20    $\pm$   0.22    &   0.238   $\pm$   0.001   &   0.31    &   9.7     $\pm$   0.7     &   296 &   4.3   \\
152.80  &   0.56    &   1   &   1   &   1.37    $\pm$   0.02    &   1.80    $\pm$   0.04    &   1.09    $\pm$   0.24    &   0.244   $\pm$   0.001   &   0.30    &   8.3     $\pm$   0.7     &   297 &   3.8   \\
154.75  &   0.67    &   1   &   1   &   2.89    $\pm$   0.03    &   3.30    $\pm$   0.07    &   2.47    $\pm$   0.41    &   0.263   $\pm$   0.001   &   0.27    &   7.4     $\pm$   1.1     &   208 &   3.6   \\
156.71  &   0.79    &   1   &   1   &   -1.85   $\pm$   0.04    &   -1.62   $\pm$   0.10    &   -2.13   $\pm$   0.63    &   0.259   $\pm$   0.002   &   0.26    &   4.1     $\pm$   1.6     &   142 &   2.5   \\

\noalign{\smallskip}
\multicolumn{12}{c}{ RZ~Vel}\\
\noalign{\smallskip}
 204.44  &   0.00    &   3   &   1   &   13.05   $\pm$   0.23    &   12.50   $\pm$   0.58    &   14.39   $\pm$   4.04    &   0.588   $\pm$   0.012   &   0.14    &   -4.4    $\pm$   4.0     &   76  &   1.3   \\
205.44  &   0.05    &   3   &   1   &   -0.47   $\pm$   0.07    &   -3.76   $\pm$   0.16    &   0.69    $\pm$   0.78    &   0.457   $\pm$   0.009   &   0.16    &   -32.0   $\pm$   2.7     &   210 &   1.5   \\
206.44  &   0.10    &   3   &   1   &   -1.90   $\pm$   0.08    &   -6.21   $\pm$   0.17    &   -0.30   $\pm$   0.96    &   0.459   $\pm$   0.016   &   0.17    &   -42.7   $\pm$   4.9     &   162 &   2.5   \\
150.49  &   0.36    &   1   &   1   &   18.65   $\pm$   0.01    &   18.46   $\pm$   0.03    &   18.76   $\pm$   0.38    &   0.219   $\pm$   0.001   &   0.38    &   -3.9    $\pm$   0.5     &   309 &   7.8   \\
152.51  &   0.46    &   1   &   1   &   28.82   $\pm$   0.01    &   28.83   $\pm$   0.02    &   28.61   $\pm$   0.60    &   0.231   $\pm$   0.001   &   0.41    &   0.8     $\pm$   0.5     &   205 &   2.8   \\
154.50  &   0.55    &   1   &   1   &   45.64   $\pm$   0.02    &   47.66   $\pm$   0.05    &   44.86   $\pm$   1.46    &   0.369   $\pm$   0.002   &   0.32    &   24.7    $\pm$   0.9     &   239 &   3.2   \\
156.49  &   0.65    &   1   &   1   &   43.14   $\pm$   0.05    &   43.91   $\pm$   0.14    &   42.87   $\pm$   2.31    &   0.526   $\pm$   0.003   &   0.23    &   6.4     $\pm$   1.1     &   178 &   1.8   \\
201.44  &   0.86    &   3   &   1   &   39.12   $\pm$   0.04    &   39.87   $\pm$   0.09    &   38.93   $\pm$   1.45    &   0.411   $\pm$   0.002   &   0.24    &   8.1     $\pm$   1.0     &   224 &   2.7   \\
202.45  &   0.90    &   3   &   1   &   29.54   $\pm$   0.06    &   29.84   $\pm$   0.15    &   29.26   $\pm$   1.09    &   0.483   $\pm$   0.003   &   0.14    &   2.8     $\pm$   1.2     &   271 &   1.2   \\
203.44  &   0.95    &   3   &   1   &   13.05   $\pm$   0.23    &   12.50   $\pm$   0.58    &   14.39   $\pm$   4.04    &   0.588   $\pm$   0.012   &   0.14    &   -4.4    $\pm$   4.0     &   76  &   1.3   \\

\hline \hline \noalign{\smallskip}
\end{tabular}
\end{center}
\end{table*}

\begin{table*}
\begin{center}
\caption[]{HARPS observations results for $\ell$~Car and RS~Pup.
See Table\ref{Tab_cepheids_results1} for legend.
\label{Tab_cepheids_results3}}
\begin{tabular}{cccccccccccc}
\hline \hline \noalign{\smallskip}

  JD$_{\mathrm c}$  &  phase   &   Cy.     &   Sp.         &  $RV_{\mathrm g}$  &   $RV_{\mathrm m}$   &   $ RV_{\mathrm c}$       &   $FWHM$   & $D$  & $A$ & $SNR$ & $\chi_{\mathrm red}^{2}$  \\
  {\tiny (a)}  &  {\tiny (b)}   &   {\tiny (c)}      &   {\tiny (d)}          &  {\tiny (e)}   &   {\tiny (f)}   &   {\tiny (g)}       &   {\tiny (h)}   & {\tiny (i)}  & {\tiny (j)} & {\tiny (k)} & {\tiny (l)}  \\

\hline

\noalign{\smallskip}
\multicolumn{12}{c}{$\ell$~Car   }\\
\noalign{\smallskip}
37.65   &   0.02    &   1   &   7   &   -13.40  $\pm$   0.01    &   -15.70  $\pm$   0.02    &   -12.19  $\pm$   0.18    &   0.416   $\pm$   0.001   &   0.26    &   -25.4  $\pm$   0.3    &   354 &   3.5   \\
40.63   &   0.10    &   1   &   5   &   -12.24  $\pm$   0.01    &   -14.22  $\pm$   0.02    &   -11.10  $\pm$   0.17    &   0.362   $\pm$   0.001   &   0.31    &   -25.2  $\pm$   0.3    &   333 &   3.6   \\
47.69   &   0.30    &   1   &   2   &   -1.57   $\pm$   0.01    &   -2.51   $\pm$   0.02    &   -0.61   $\pm$   0.24    &   0.274   $\pm$   0.001   &   0.42    &   -16.4  $\pm$   0.4    &   275 &   7.5   \\
48.62   &   0.33    &   1   &   2   &   -0.01   $\pm$   0.01    &   -0.76   $\pm$   0.01    &   0.59    $\pm$   0.10    &   0.257   $\pm$   0.001   &   0.40    &   -13.7  $\pm$   0.2    &   438 &   10.3  \\
49.67   &   0.36    &   1   &   2   &   1.79    $\pm$   0.01    &   1.09    $\pm$   0.02    &   2.34    $\pm$   0.13    &   0.260   $\pm$   0.001   &   0.40    &   -12.7  $\pm$   0.3    &   374 &   6.0   \\
15.72   &   0.40    &   1   &   3   &   4.92    $\pm$   0.00    &   4.33    $\pm$   0.01    &   5.42    $\pm$   0.12    &   0.276   $\pm$   0.001   &   0.40    &   -10.2  $\pm$   0.2    &   405 &   7.0   \\
51.68   &   0.41    &   2   &   4   &   5.23    $\pm$   0.00    &   4.62    $\pm$   0.01    &   5.75    $\pm$   0.13    &   0.274   $\pm$   0.001   &   0.39    &   -10.4  $\pm$   0.2    &   352 &   6.0   \\
16.69   &   0.43    &   1   &   3   &   6.64    $\pm$   0.01    &   6.13    $\pm$   0.02    &   7.15    $\pm$   0.21    &   0.289   $\pm$   0.001   &   0.39    &   -8.3   $\pm$   0.3    &   293 &   5.6   \\
52.64   &   0.44    &   2   &   2   &   6.93    $\pm$   0.01    &   6.30    $\pm$   0.02    &   7.42    $\pm$   0.18    &   0.285   $\pm$   0.001   &   0.39    &   -10.3  $\pm$   0.3    &   376 &   5.4   \\
17.71   &   0.46    &   1   &   3   &   8.46    $\pm$   0.01    &   8.11    $\pm$   0.01    &   8.86    $\pm$   0.15    &   0.300   $\pm$   0.001   &   0.38    &   -5.5   $\pm$   0.2    &   444 &   8.7   \\
53.69   &   0.47    &   2   &   2   &   8.70    $\pm$   0.01    &   8.31    $\pm$   0.02    &   9.00    $\pm$   0.18    &   0.288   $\pm$   0.001   &   0.37    &   -6.3   $\pm$   0.3    &   390 &   5.0   \\
54.67   &   0.50    &   2   &   2   &   10.36   $\pm$   0.01    &   10.18   $\pm$   0.02    &   10.52   $\pm$   0.22    &   0.296   $\pm$   0.001   &   0.35    &   -2.9   $\pm$   0.3    &   340 &   4.7   \\
55.70   &   0.53    &   2   &   2   &   11.99   $\pm$   0.01    &   12.04   $\pm$   0.02    &   12.08   $\pm$   0.20    &   0.316   $\pm$   0.001   &   0.35    &   0.6    $\pm$   0.2    &   418 &   4.5   \\
20.84   &   0.54    &   1   &   3   &   14.11   $\pm$   0.01    &   14.65   $\pm$   0.03    &   13.93   $\pm$   0.35    &   0.335   $\pm$   0.001   &   0.33    &   7.4    $\pm$   0.4    &   218 &   2.2   \\
56.70   &   0.55    &   2   &   2   &   13.46   $\pm$   0.01    &   13.75   $\pm$   0.02    &   13.42   $\pm$   0.25    &   0.331   $\pm$   0.001   &   0.34    &   4.0    $\pm$   0.3    &   378 &   4.2   \\
21.85   &   0.57    &   1   &   3   &   15.01   $\pm$   0.01    &   15.66   $\pm$   0.02    &   14.76   $\pm$   0.22    &   0.358   $\pm$   0.001   &   0.32    &   8.3    $\pm$   0.2    &   379 &   3.6   \\
57.70   &   0.58    &   2   &   2   &   14.82   $\pm$   0.01    &   15.35   $\pm$   0.02    &   14.67   $\pm$   0.27    &   0.347   $\pm$   0.001   &   0.34    &   7.0    $\pm$   0.3    &   384 &   3.7   \\
58.71   &   0.61    &   2   &   2   &   16.00   $\pm$   0.01    &   16.83   $\pm$   0.02    &   15.70   $\pm$   0.27    &   0.365   $\pm$   0.001   &   0.33    &   10.5   $\pm$   0.3    &   412 &   5.3   \\
23.66   &   0.62    &   1   &   4   &   16.98   $\pm$   0.01    &   17.92   $\pm$   0.02    &   16.55   $\pm$   0.20    &   0.388   $\pm$   0.001   &   0.31    &   11.1   $\pm$   0.2    &   428 &   3.5   \\
24.85   &   0.66    &   1   &   4   &   17.94   $\pm$   0.01    &   19.15   $\pm$   0.02    &   17.43   $\pm$   0.22    &   0.389   $\pm$   0.001   &   0.30    &   14.0   $\pm$   0.2    &   374 &   3.6   \\
25.87   &   0.69    &   1   &   4   &   18.52   $\pm$   0.01    &   19.63   $\pm$   0.02    &   18.00   $\pm$   0.19    &   0.404   $\pm$   0.001   &   0.30    &   12.4   $\pm$   0.2    &   492 &   4.2   \\
26.85   &   0.71    &   1   &   5   &   19.03   $\pm$   0.01    &   20.27   $\pm$   0.02    &   18.50   $\pm$   0.24    &   0.426   $\pm$   0.001   &   0.31    &   13.4   $\pm$   0.2    &   436 &   5.9   \\
28.69   &   0.77    &   1   &   5   &   20.12   $\pm$   0.01    &   21.26   $\pm$   0.02    &   19.50   $\pm$   0.22    &   0.426   $\pm$   0.001   &   0.29    &   12.2   $\pm$   0.2    &   433 &   3.1   \\
29.65   &   0.79    &   1   &   5   &   20.33   $\pm$   0.01    &   21.61   $\pm$   0.02    &   19.63   $\pm$   0.23    &   0.445   $\pm$   0.001   &   0.29    &   13.2   $\pm$   0.2    &   439 &   4.7   \\
30.80   &   0.83    &   1   &   7   &   19.76   $\pm$   0.01    &   20.83   $\pm$   0.01    &   19.05   $\pm$   0.19    &   0.450   $\pm$   0.001   &   0.28    &   11.0   $\pm$   0.1    &   446 &   6.6   \\
31.66   &   0.85    &   1   &   5   &   17.77   $\pm$   0.01    &   18.65   $\pm$   0.02    &   17.15   $\pm$   0.24    &   0.458   $\pm$   0.001   &   0.27    &   8.9    $\pm$   0.2    &   391 &   5.3   \\
32.72   &   0.88    &   1   &   5   &   11.81   $\pm$   0.01    &   12.17   $\pm$   0.02    &   11.51   $\pm$   0.20    &   0.481   $\pm$   0.001   &   0.24    &   3.4    $\pm$   0.2    &   431 &   3.8   \\
33.63   &   0.91    &   1   &   5   &   3.92    $\pm$   0.01    &   3.06    $\pm$   0.03    &   4.31    $\pm$   0.20    &   0.483   $\pm$   0.001   &   0.23    &   -8.2   $\pm$   0.2    &   371 &   4.1   \\
34.67   &   0.93    &   1   &   5   &   -4.93   $\pm$   0.01    &   -6.89   $\pm$   0.02    &   -3.99   $\pm$   0.15    &   0.460   $\pm$   0.001   &   0.24    &   -19.5  $\pm$   0.2    &   486 &   8.3   \\
35.66   &   0.96    &   1   &   7   &   -9.85   $\pm$   0.01    &   -11.93  $\pm$   0.02    &   -8.81   $\pm$   0.14    &   0.434   $\pm$   0.001   &   0.24    &   -21.8  $\pm$   0.2    &   421 &   6.0   \\
36.65   &   0.99    &   1   &   6   &   -12.39  $\pm$   0.01    &   -14.56  $\pm$   0.02    &   -11.29  $\pm$   0.21    &   0.423   $\pm$   0.001   &   0.24    &   -23.4  $\pm$   0.3    &   327 &   2.8   \\

\noalign{\smallskip}
\multicolumn{12}{c}{RS~Pup  }\\
\noalign{\smallskip}

56.68   &   0.02    &   1   &   1   &   3.58    $\pm$   0.03    &   1.58    $\pm$   0.07    &   4.33    $\pm$   0.41    &   0.433   $\pm$   0.003   &   0.20    &   -20.1  $\pm$   0.9    &   347 &   2.1   \\
58.69   &   0.07    &   1   &   1   &   5.31    $\pm$   0.03    &   3.78    $\pm$   0.07    &   5.94    $\pm$   0.46    &   0.395   $\pm$   0.002   &   0.23    &   -17.1  $\pm$   0.9    &   277 &   2.0   \\
60.68   &   0.12    &   1   &   1   &   7.95    $\pm$   0.02    &   6.97    $\pm$   0.05    &   8.50    $\pm$   0.42    &   0.370   $\pm$   0.001   &   0.26    &   -11.0  $\pm$   0.7    &   315 &   2.3   \\
62.67   &   0.17    &   1   &   1   &   10.99   $\pm$   0.02    &   10.28   $\pm$   0.05    &   11.48   $\pm$   0.55    &   0.343   $\pm$   0.001   &   0.29    &   -8.5   $\pm$   0.7    &   258 &   2.8   \\
64.68   &   0.22    &   1   &   1   &   14.21   $\pm$   0.01    &   13.70   $\pm$   0.04    &   14.55   $\pm$   0.41    &   0.307   $\pm$   0.001   &   0.32    &   -6.8   $\pm$   0.5    &   328 &   3.2   \\
66.66   &   0.26    &   1   &   1   &   17.46   $\pm$   0.01    &   16.96   $\pm$   0.03    &   17.96   $\pm$   0.58    &   0.296   $\pm$   0.001   &   0.36    &   -6.9   $\pm$   0.5    &   308 &   3.8   \\
150.48  &   0.28    &   3   &   1   &   18.43   $\pm$   0.01    &   17.89   $\pm$   0.03    &   18.96   $\pm$   0.64    &   0.317   $\pm$   0.001   &   0.37    &   -6.5   $\pm$   0.5    &   310 &   2.9   \\
152.49  &   0.33    &   3   &   1   &   21.56   $\pm$   0.02    &   21.15   $\pm$   0.04    &   21.79   $\pm$   0.59    &   0.269   $\pm$   0.001   &   0.34    &   -6.9   $\pm$   0.7    &   249 &   7.2   \\
154.49  &   0.38    &   3   &   1   &   24.80   $\pm$   0.02    &   24.38   $\pm$   0.05    &   25.12   $\pm$   0.89    &   0.271   $\pm$   0.001   &   0.36    &   -6.9   $\pm$   0.7    &   211 &   3.0   \\
156.48  &   0.43    &   3   &   2   &   27.90   $\pm$   0.01    &   27.52   $\pm$   0.04    &   28.12   $\pm$   0.67    &   0.264   $\pm$   0.001   &   0.34    &   -6.5   $\pm$   0.6    &   202 &   4.1   \\
48.61   &   0.83    &   1   &   1   &   47.56   $\pm$   0.04    &   50.66   $\pm$   0.09    &   46.52   $\pm$   1.69    &   0.499   $\pm$   0.004   &   0.24    &   27.3   $\pm$   1.1    &   239 &   3.3   \\
51.64   &   0.90    &   1   &   1   &   30.44   $\pm$   0.11    &   25.29   $\pm$   0.27    &   31.52   $\pm$   2.02    &   0.955   $\pm$   0.010   &   0.12    &   -22.6  $\pm$   1.6    &   249 &   2.6   \\
52.63   &   0.93    &   1   &   1   &   16.40   $\pm$   0.12    &   12.85   $\pm$   0.30    &   18.39   $\pm$   1.97    &   0.681   $\pm$   0.012   &   0.13    &   -22.8  $\pm$   2.5    &   173 &   1.3   \\
54.66   &   0.97    &   1   &   1   &   5.26    $\pm$   0.05    &   2.97    $\pm$   0.13    &   6.07    $\pm$   0.71    &   0.479   $\pm$   0.005   &   0.17    &   -20.9  $\pm$   1.5    &   247 &   1.0   \\

\hline \hline \noalign{\smallskip}
\end{tabular}
\end{center}
\end{table*}
\end{document}